\documentclass[aps,prf, reprint, amssymb,groupedaddress,longbibliography,onecolumn]{revtex4-2}

\usepackage[colorlinks=true,citecolor=blue,linkcolor=blue]{hyperref}
\usepackage{natbib}
\usepackage{outlines}

\usepackage{indentfirst}
\usepackage{amsmath}
\usepackage{xfrac}
\usepackage{subfigure}

\usepackage[usenames,dvipsnames]{pstricks}
\usepackage{epstopdf}

\usepackage{caption}
\usepackage{sidecap}
\usepackage{mathrsfs}

\usepackage[english]{babel}
\usepackage[utf8]{inputenc}
\usepackage{graphicx}
\usepackage{caption}
\usepackage{natbib}
\usepackage{fixltx2e}
\usepackage{textgreek}
\usepackage{textcomp}
\usepackage{multirow}
\usepackage{array}
\usepackage{pbox}
\usepackage{siunitx}
\usepackage{xcolor}
\usepackage{float}
\definecolor{cinnamon}{rgb}{0.82, 0.41, 0.12}
\definecolor{pink}{rgb}{0.858, 0.188, 0.478}
\definecolor{black}{rgb}{0.0, 0.0, 0.0}
\usepackage{gensymb}

\usepackage{mathtools}
\usepackage{amsmath}
\usepackage{physics}

\usepackage{tikz}
\usepackage{color}	
\usetikzlibrary{positioning,shapes,arrows}
\usetikzlibrary{shapes.symbols,patterns}
\tikzstyle{line} = [draw, -latex']
\tikzstyle{block} = [rectangle, draw, text width=9em, text centered, rounded corners, minimum height=2em]
\usepackage{mathtools,etoolbox}

\definecolor{lblue1}{RGB}{234,243,250}
\definecolor{lgreen1}{RGB}{200,245,213}
\definecolor{lorange1}{RGB}{255,219,179}
\definecolor{myindigo}{RGB}{74,0,125}
\usepackage[normalem]{ulem}

\DeclareCaptionJustification{justified}{\justifying}
\usepackage{titlesec}
\titlelabel{\thetitle.\ }

\begin{document}

\title{Formation of colloidal threads in geometrically varying flow-focusing channels}
 
\author{V$.$ Krishne Gowda$^1$}\email[]{krivi@mech.kth.se}
\author{Cecilia Rydefalk$^{1}$, L$.$ Daniel S\"{o}derberg$^{2,3}$}
\author{Fredrik Lundell$^{1,3}$}
\affiliation{$^1$FLOW, Dept$.$ of  Engineering Mechanics, KTH Royal Institute of Technology, SE-100 44 Stockholm, Sweden\\$^2$Dept$.$ of  Fibre and Polymer Technology, KTH Royal Institute of Technology, SE-100 44 Stockholm, Sweden\\$^3$Wallenberg Wood Science Center, KTH Royal Institute of Technology, SE-100 44 Stockholm, Sweden}


\begin{abstract}
 When two miscible fluids are brought into contact with each other, the concentration gradients induce stresses. These are referred to as Korteweg stresses and are analogous to interfacial tension between two immiscible fluids, thereby acting as an effective interfacial tension (EIT) in inhomogeneous miscible fluid pairs. Effective interfacial tension governs the formation of a viscous thread in flow-focusing of two miscible fluids. To further investigate its significance, we have studied thread formation of a  percolated colloidal dispersion focused by its own solvent.  Experiments are combined with three-dimensional (3-D) numerical methods to systematically expand previous knowledge utilising different flow-focusing channel setups. In the reference flow-focusing configuration, the sheath flows impinge the core flow orthogonally while in four other channel configurations, the sheath flows impinge the core flow at an oblique angle that is both positive and negative with respect to the reference sheath direction.  As an initial estimate of the EIT between the colloidal dispersion-solvent system, we fit the experimentally determined thread shape in the reference configuration to a master curve that depends on EIT through an effective capillary number. By numerically reproducing these experimental results, it is concluded that the estimated EIT is within 25\% of the “optimal”  EIT value that can be deduced by iteratively fitting the numerical results to the experimental measurements. Regardless of channel configurations, further numerical calculations performed using the optimal EIT evaluated from the reference configuration show good agreement with the experimental findings in terms of 3-D thread shapes, wetted region morphologies, and velocity flow fields. Through an additional numerical study, we demonstrate that the 3-D flow characteristics observed experimentally in oblique channel configurations can be precisely replicated with orthogonal sheath flows if the sheath channel width is adjusted. The one-to-one comparison and analysis of numerical and experimental findings unveil the crucial role of EIT on the thread detachment from the top and bottom walls of the channel, bringing useful insights to understand the physical phenomenons involved in miscible systems with a high-viscosity contrast. These key results provide the foundation to tune the flow-focusing for specific applications, for example in tailoring the assembly of nanostructured materials.
 
\end{abstract}

\maketitle

\begin{centering}
\section{Introduction}
\label{intro}
\end{centering}

A boundary between two fluid phases is identifiable, irrespective of whether the fluids are immiscible or miscible. In the case of immiscible fluids, the boundary zone is clearly distinguished through a distinct interface created by the equilibrium interfacial tension acting between the two fluids. On the other hand, in the context of miscible fluids, no such distinct interface exists as the two fluids are fully mixed at equilibrium. However, when two miscible fluids come into contact, a boundary between them can persist till the time the two fluids eventually form an equilibrated homogeneous mixture due to diffusive mixing. This boundary zone can act  as a \emph{de-facto} interface with some of its properties resembling that of a distinct interface observed between immiscible fluids~\cite{joseph1990fluid,garik1991,joseph1993fundamentals,anderson1998,atencia2004}. In 1901,  \citet{Korteweg1901} first proposed that when two miscible fluids are brought into contact,  the composition inhomogeneties/gradients of the fluid property at the zone of contact gives rise to additional stresses (so-called Korteweg stresses).  These stresses effectively mimic capillary-like stress effects across the boundary zone that can be seen as a sharp \emph{de-facto} interface. Accordingly, analogous to the interfacial tension $\gamma$ in the immiscible fluids, an effective interfacial tension (EIT) for  miscible fluids in non-equilibrium state, commonly denoted as $\Gamma_{e}$, can be written as:
\begin{equation} \label{composition} \Gamma_{e} = K \frac{\Delta\Phi^{2}}{\delta},\end{equation} 
where $K$ is the Korteweg factor accounting for the relevant interaction effects (e.g.~particle-solvent interactions in miscible complex fluids) between the two fluids~\cite{truzzolillo2014,truzzolillo2016}, $\delta$ is the interface thickness and $\Delta \Phi$ is the variation in composition or volume fraction $\Phi$. As  can be seen from Eq.~(\ref{composition}), $\Gamma_{e}$ exists  as long as the composition gradients persist at the \emph{de-facto} interface, and as the interface smears out due to diffusion over time, the EIT  goes to zero.

 The role of these transient stresses or EIT in non-equilibrium miscible fluids is evident in numerous multi-fluid dynamical processes occuring at short times. Examples are  microgravity experiments~\cite{bessonov2005}, evolution of miscible droplets~\cite{chen2001,chen2002},  modeling of hydrodynamic instabilities like viscous fingering or in Hele-Shaw flows~\cite{chen2008,dias2013},  stabilization of Rayleigh-Taylor instabilities induced by evaporation between a polymer solution and its own solvent~\cite{mossige2020} and so on. Some of the recent experimental techniques explored to measure the EIT between miscible fluids are through the evolution of drop shape~\cite{pojman2006, carbonaro2020}, examination of hydrodynamic instabilities~\cite{truzzolillo2014, shevtsova2016}, and probing of capillary waves by light scattering~\cite{may1991,cicuta2001}. In spite of these attempts, measuring the EIT between miscible fluids is intrinsically difficult  due to ultra-low values ($\Gamma_{e}$~$\sim$~$10^{-4}$$-$$10^{1}$ mN~m$^{-1}$) and absence of a distinct interface~\citep{truzzolillo2014,truzzolillo2016, truzzolillo2017}.

Lately, in the framework of Korteweg’s theory for miscible fluids, employing a microfluidic flow-focusing setup, we have proposed a methodology  with a possibility to determine the EIT between two miscible fluids. In short, $\Gamma_{e}$ can be estimated by measuring the spatial evolution of the thread shape formed by these two fluids, and equating it with a master curve obtained based on a  simple  scaling model~\cite{gowda2019effective}. However,  this method was not yet fully exploited. 
 
\begin{figure} [tbp!]
	\begin{centering}
		\includegraphics[width=1\textwidth]{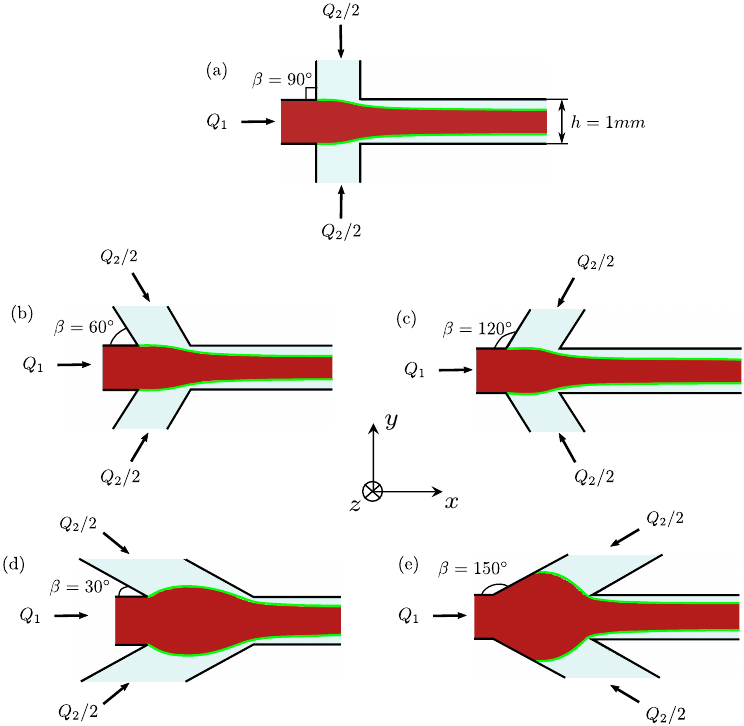}
		\caption{(a) - (e) Schematic illustration of the top view of geometrically varying flow-focusing channel configurations with various confluence angles~$\beta$.  The red color represents the core fluid entering the central inlet channel with a volumetric flow rate~$Q_{1}$.  The light cyan color denotes the sheath fluid entering from the side inlet channels at an angle~$\beta$ with a flow rate of~$Q_{2}/2$ in each channel. The central channel has a square cross-section with sidelength $h$.}
		\label{fig:fig1}
	\end{centering}
\end{figure}

 Flow-focusing essentially comprises of a central core fluid focused by two side (sheath) fluids as shown schematically in  Fig.~\ref{fig:fig1}(a). For miscible systems, the Péclet number $Pe=Uh/D$,  which measures the relative importance of convective and diffusive effects, is the  relevant non-dimensional quantity to describe the flow with an average velocity~$U$ in a microchannel of width $h$, and $D$ being the diffusion coefficient between the two fluids~\cite{atencia2004,cubaud2006}. When a pair of miscible fluids is passed through the flow-focusing channel, the Péclet number dictates the flow-patterns. At high Péclet number under laminar flow conditions, where diffusion is almost negligible, viscous threads are formed, while at low Péclet number, the threads undergo diffusive instabilities leading to a wide diversity of flow-patterns~\cite{cubaud2012,bonhomme2012, cubaud2014}. It is even observed that the miscible threads at high Péclet number is equivalent to the multi-fluid viscous flow problem, and could be employed to examine and realise the role of diffusion~\cite{cubaud2009,cubaud2020}. 

For a high Péclet number system, with a pair of miscible fluids comprising of a colloidal dispersion and its own solvent, we have detected and reported that the characteristics of spatial thread evolution could be accurately captured and modelled as an  immiscible fluid problem with a very weak interfacial tension $\gamma$ for a set of flow rates and specified rheology of the two fluids~\cite{gowda2019effective}.  Ideally, at high Péclet number, the time scale for interdiffusion between the colloidal dispersion and its own solvent is almost negligible compared to the convective time scale of the fluids in the channel. In such a scenario, the presence of a sharp \emph{de-facto} interface due to composition gradients is expected in the experiments~\cite{atencia2004,truzzolillo2014, truzzolillo2016}. However, it is extremely difficult to access such a sharp \emph{de-facto} interface between the two fluids experimentally.  The miscible viscous thread structures formed at high Péclet number are in an out-of-equilibrium  state and occur before the two fluids are fully mixed. 

In such occurences,  it is possible to reckon the \emph{de-facto} interface  by appyling a weak interfacial tension $\gamma$ in numerical modelling. In reality, the weak interfacial tension $\gamma$ accounts for the Korteweg stresses induced by the composition gradients in the experiments.  The experimental observations were reproduced numerically with a very weak interfacial tension $\gamma$ ($\mathcal{O}$$\sim$$10^{-2}$ mN~m$^{-1}$). Moreover, the magnitude of  $\gamma$ corroborated with the experimental measurements of  previous studies conducted for miscible complex fluids involving colloidal dispersions or  polymer solutions and its own solvent~\cite{truzzolillo2016, truzzolillo2017}. Thus, it was established that the very weak interfacial tension $\gamma$ acts as  an effective interfacial tension  in the experiments  with $\gamma$~$\equiv$~$\Gamma_{e}$.

Furthermore, in microfluidic systems,  the interfacial tension and viscous dissipation of energy dominate over  inertial effects~\cite{anna2003,garstecki2004,whitesides2006} due to the small length scales. Pertaining to miscible systems, and in the context of weakly diffusive threads, the phenomenon of viscous dissipation of energy is pertinent as the viscosity contrast between the fluid pairs is fairly large~\cite{cubaud2009,cubaud2012}.  In addition to these effects, we noted that the effect of EIT due to concentration gradients also play a significant role in such systems at small scales~\cite{gowda2019effective}. For a finite length of the channel, the streamwise spatial evolution of the high-viscosity thread cross-section from non-circular (near-ellipsoidal) to a circular shape was found to be dependent on EIT. Indeed, it is plausible for  a  high-viscosity non-circular thread to evolve naturally and adapt a circular shape through minimization of  energy via viscous dissipation. But, such a natural process was found to be inconsequential compared to  the effect of EIT. This was evident from the estimation of time and length scales derived from a simple scaling model~\cite{gowda2019effective}, wherein, the spatial evolution of  the high-viscosity thread from near-ellipsoidal cross-section to circular shape could be quantified by an effective capillary number.  Analogous to the capillary number in immiscible systems, the effective capillary number based on  $\Gamma_{e}$ for the non-equilibirum miscible system could be  defined as,
\begin{equation} \label{effective capillary}
Ca_e=\frac{\eta Q}{h^2\Gamma_{e}}
\end{equation}

\noindent where $\eta$ is the dynamic viscosity and $Q$ is the flow rate of the core fluid thread, and $h$ is the width of channel.  Thus, when the streamwise spatial coordinate is normalised with the above $Ca_{e}$,  all the spatial evolution of  the thread heights at different  $\gamma$ ($\Gamma_{e}$) collapse on to a master curve. 

In the present work, we widen the studies of viscous thread formation at high Péclet number in miscible environments into a systematic investigation employing geometrically varying flow-focusing setups with a confluence angle~$\beta$ as illustrated in Fig.~\ref{fig:fig1}. The \emph{Confluence angle}~$\beta$ refers to the angle made by the side (sheath) flow channel inlets with the central (core) flow channel inlet. We characterise the flow in three dimensions (3-D) both experimentally and numerically employing optical coherence tomography (OCT) and volume of  fluid (VoF) method~\cite{hirt1981volume} implemented in the open-source code OpenFOAM~\cite{OpenCFD}. 

There are three profound objectives for this study. The first is to investigate the effect of confluence angles on the 3-D shape of  the thread structures. Here, we aim to understand the influence of sheath fluid impingement and geometrical channel effects on the thread formation and wetted region morphologies. Through this, we address an important aspect: whether the process of thread detachment from the top and bottom walls of the channel in miscible systems~\cite{cubaud2009, cubaud2014} occurs (i) naturally through the self-lubrication principle~\cite{cubaud2012}, a phenomenon associated with the effect of viscous dissipation of energy and originally observed in core-annular flow involving high-viscosity contrast immiscible fluid pairs~\cite{joseph1984, joseph1993fundamentals}, or  (ii) through some other mechanisms  with respect to high-viscosity contrast miscible fluid pairs.  The effects of confluence angle~$\beta$ could also be potentially useful to understand how the sheath flow momentum affects the system, and identify the means to achieve efficient extensional flows.  The second objective is, to exploit the experimentally measured 3-D spatial evolution of the thread shape together with an effective capillary number dependent master curve to estimate an experimentally intractable variable, namely the $\Gamma_{e}$  between two miscible fluids. The final goal is, to bring out the implications of EIT in microfluidic flow systems. In addition to these, a valuable feature of microfluidic flows is demonstrated, where the 3-D flow characteristics of different confluence angle geometries can be replicated with a  geometry of  $\beta$ = 90$\degree$ by altering  the sheath flow channel  widths. The present study also illustrates how additional insights into the physical mechanisms acting experimentally in the complex microfluidic systems can be gained from a diligent analysis of numerical calculations. These insights are difficult to ascertain in experiments due to inherent composition-dependent fluid properties.

As the experimental  fluids, similar to our previous work~\cite{gowda2019effective}, we use  a colloidal dispersion for the core and its own solvent as the sheath fluid. The ingredients of  the colloidal dispersion consists of cellulose nanofibrils (CNF) dispersed in water. Such nanofibrils have been assembled into high-performance structural cellulose filaments via hydrodynamic  focusing~\cite{haakansson2014hydrodynamic, mittal2018}.  Understanding the underlying flow behaviour among various geometrical configurations is critical for controlling the hydrodynamic assembly~\cite{Brouzet_2018,Brouzet_2019}. Moreover, these colloidal systems exhibit variant composition-dependent fluid properties based on particle structure and interparticle interactions, and the $\Gamma_{e}$ for such colloidal dispersion and its solvent system is expected to vary considerably ~\cite{truzzolillo2015,truzzolillo2016}.

The organisation of  the paper is as follows.  In Sec.~\ref{methods}, we give an overview of the experimental setup and a brief outline of the numerical method.  In Sec.~\ref{thread shapes}, we compare and discuss the numerical and experimental 3-D thread topologies and wetted region morphologies on the top and bottom walls for various flow-focusing configurations emphasizing the role of confluence angle $\beta$. In Sec.~\ref{model curve}, we briefly recall the scaling model, and the master curve, and use it to estimate the $\Gamma_{e}$ for the present experimental fluids. The estimated $\Gamma_{e}$, in turn, is verified with the value of $\Gamma_{e}$ obtained through an optimization procedure.  In Sec.~\ref{centreline velocities}, numerical and experimental results of the centreline velocity is compared. Further, a numerical comparison of strain rates along the centreline in different geometrical configurations is undertaken to dissect the effects of confluence angle~$\beta$ on the flow-field. In Sec.~\ref{cross-sectional velocities}, we present numerical results of cross-sectional velocity distributions  for all the confluence angle geometries. In Sec.~\ref{duplicate}, replicability of the 3-D flow features of different confluence angles~$\beta$  in modified flow-focusing channels with $\beta$~=~90$\degree$ is presented. In Sec.~\ref{implication}, a remark  highlighting the significance of  EIT in microfluidic channels is elucidated, and finally a brief summary of  the conclusions  is provided in Sec.~\ref{conclusions}.  \\

\begin{centering}\section{Experimental and numerical setups}\label{methods}\end{centering}

\begin{centering}\subsection{Experimental setup}\label{exp-setup}\end{centering}

\begin{centering}\subsubsection{Flow-focusing geometries}\label{geom-setup}\end{centering}

In this work, we employ five distinct types of flow-focusing channel setups. The setups are planar with square cross-sections of sidelength $h = 1$~mm and have the geometrical configurations as shown in Fig.~\ref{fig:fig1}. All the geometries have one main central inlet channel for the core flow and two sheath flow inlets inclined with  a  confluence angle~$\beta$.  The confluence angle is varied between 30$\degree$ and 150$\degree$ with an interval of  30$\degree$, so as to systematically investigate the impact of sheath flow impingement and channel effect on the core fluid 3-D thread topology and flow-field. 

All the channel geometries are built using a stainless-steel plate of $1$~mm thickness similar to our previous works involving flow-focusing configuration with confluence angle~$\beta$~=~90$\degree$~\cite{haakansson2014hydrodynamic,Brouzet_2018,Brouzet_2019,gowda2019effective}. The steel channel plate is enclosed on both sides with layers of  aluminum plates and a cyclic olefin copolymer (COC) film forming an assembly of  `aluminum plate$-$COC$-$steel channel$-$COC$-$aluminum plate' sandwich. The fluids are injected at constant volumetric flow rates into the core and sheath flow channel inlets by two syringe pumps (WPI, AI-$4000$). Furthermore, all the geometrical configurations (Figs.~\ref{fig:fig1}(b),~\ref{fig:fig1}(c) and  \ref{fig:fig1}(d),~\ref{fig:fig1}(e)) will be discussed in relative to the reference configuration (Fig.~\ref{fig:fig1}(a)), and this choice will be clarified in Sec.~\ref{thread shapes}.

\begin{figure} [tbp!]
	\begin{centering}
		\includegraphics[width=0.7\textwidth]{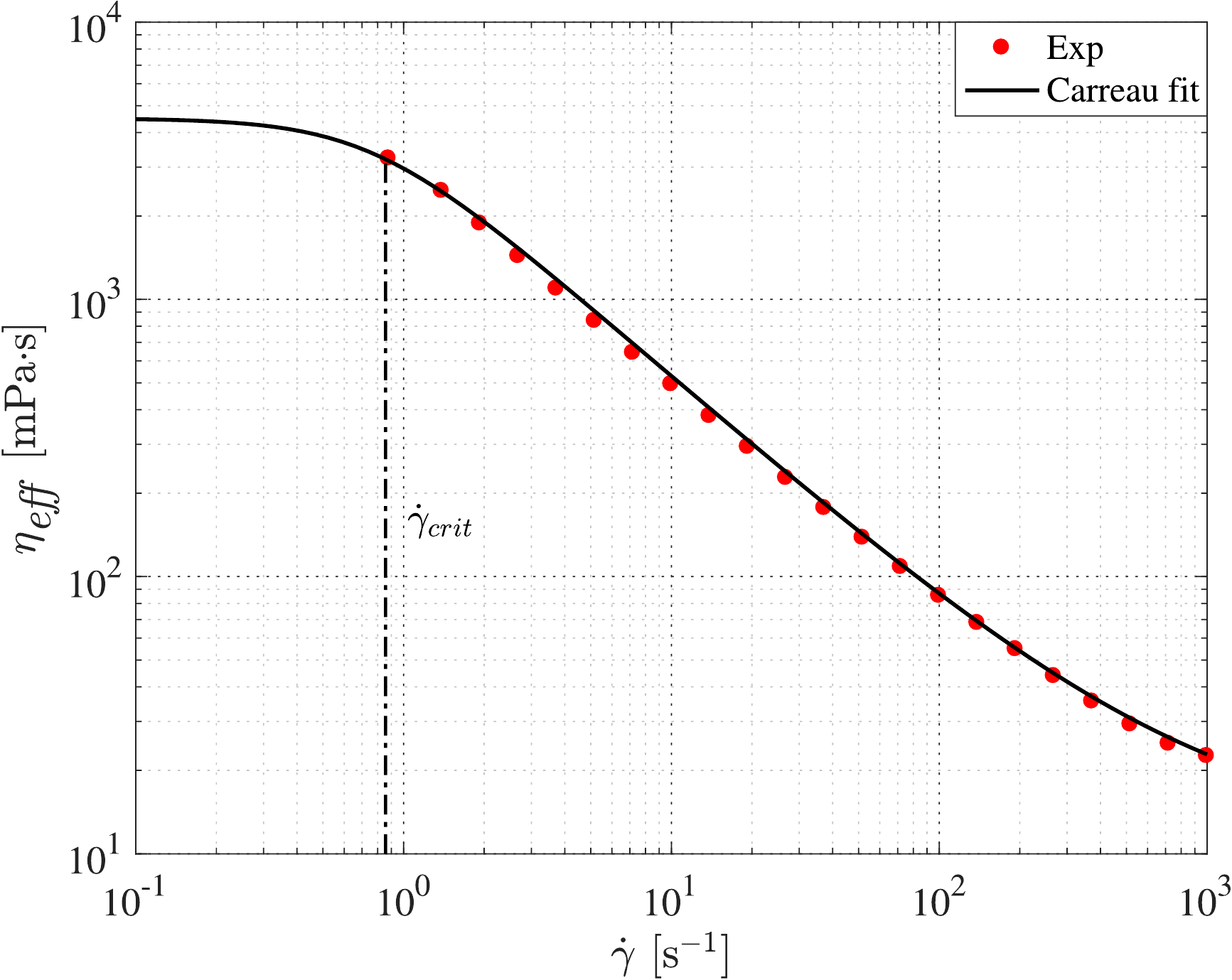} 		
		\caption{ Shear viscosity measurements of the colloidal dispersion represented by red dots. The solid black line represents the non-Newtonian Carreau model fit (see Eq.~\ref{Car}), and the vertical dashed-dotted line denotes the critical shear rate $\dot{\gamma}_{crit}$ for the dispersion. }\label{Rheology}
		\label{fig:fig2}    
	\end{centering}
\end{figure}

In all the configurations, the experimental measurements are performed at constant volumetric flow rates (see Fig.~\ref{fig:fig1}) with $Q_{1}$ = 6.5  mm$^{3}~$s$^{-1}$ and $Q_{2}$~=~ $7.5$~mm$^{3}~$s$^{-1}$, respectively. The core fluid is a colloidal dispersion exhibiting a non-Newtonian viscosity behaviour as shown in Fig.~\ref{fig:fig2}. The sheath fluid is a de-ionized (DI) water of viscosity $\eta_2=1$~mPa$~$s. \\

\begin{centering}\subsubsection{Dispersion material and its rheological properties}\label{fluids}\end{centering}

The colloidal dispersion is composed of cellulose nanofibrils (CNF) suspended in water.  Cellulose nanofibrils were obtained by liberating fibrils from never dried sulfite softwood pulp (Domsjö Fabriker AB, Sweden) supplied by RISE (Research Institute of Sweden). Before defibrillation, never dried sulfite softwood pulp was subjected to TEMPO-mediated oxidation following the protocol described elsewhere~\cite{saito2007,isogai2011tempo}. Thereafter, defibrillation of the oxidized pulp was carried out by passing through a high pressure ({$1600$} {bars}) Microfluidizer (M-$110$EH, microfluidics) with  $400/200~\mu$m (1 pass) and $200/100~\mu$m (4 passes) wide chambers connected in series. The resulting output is a cellulose nanofibril dispersion of 1~wt\%. Finally, a transparent colloidal dispersion of concentration $3$~g$~$dm$^{-3}$  was obtained by further dilution from  1~wt\% to 0.3~wt\% and homogenization by an Ultra-Turrax dispersing tool (IKA, Sweden) for {10} {min} at {12000} revolutions per minute.  The typical fibril lengths~$L$ vary from $100$$-$$1500$~nm, and the average fibril diameter~$d$ is $2.3\pm0.7$~nm as determined by Atomic Force Microscopy (AFM, Dimension 3100 SPM, Veeco, USA) measurements. 

The rheological characterization of the colloidal dispersion was performed using a bob and cup Kinexus pro+ rheometer (Malvern). This rheometer is well suited for accurate measurements of viscosity (indicated by red dots) over a range of shear rates, as observed in Fig.~\ref{fig:fig2}. The dispersion  displays a non-Newtonian shear thinning behaviour~\cite{martoia2016, Nechyporchuketal2016, geng2018}. 

 The rheological data can be described well by a Carreau model

\begin{centering} 
	\begin{equation}
	\eta_{\textrm{\textit{eff}}}=\eta_{\textrm{\textit{inf}}}+(\eta_{0}-\eta_{\textrm{\textit{inf}}})[1+(\tau\dot{\gamma})^2]^\frac{n-1}{2},
	\label{Car}
	\end{equation}
\end{centering}

\noindent where $\eta_{\textrm{\textit{eff}}}$ is the shear viscosity, $\eta_{\textrm{\textit{inf}}}$ the infinite shear viscosity, $\eta_{0}$ the zero shear viscosity,  $\tau$ the relaxation time,  $\dot{\gamma}$~the shear rate,  and $\mathit{n}$ the power index. The parameters of the  Carreau model fit denoted by the solid black line in Fig.~\ref{Rheology} are as follows: $\eta_{\textrm{\textit{inf}}}=12 $~mPa$~$s, $\eta_0=4500 $~mPa$~$s, $\tau=1.306$~s, and $n=0.16$. 

The shear thinning behaviour is the consequence of micro-structural re-arrangements  observed in the colloidal dispersion resulting from fibril/solvent molecular interactions, electro-static interactions or due to effects of  Brownian motion and fibre entanglement~\cite{switzer2003}. As the dimensions of the fibrils are very small, Brownian motion along  with the entanglement effects is  dominant over other interactions. At low shear rates, the micro-structure of the fibrils are in disordered or isotropic arrangement. Once the shear rates are high enough to overcome Brownian effects, fibrils tend to reorganize by aligning and orienting towards the flow direction. As a result, the viscosity starts to decrease at a particular critical shear rate $\dot{\gamma}_{crit}$ indicating the onset of shear thinning. The critical shear rate can be obtained by taking the  inverse of the orientational relaxation time of fibrils, $\dot{\gamma}_{crit}$ = ${\tau}_{r}^{-1}$ where ${\tau}_{r}$~=$1/(6{D}_{r}$)~\cite{doi1988}. The orientational relaxation time of a Brownian system can be estimated by calculating the rotational diffusion coefficient~$D_{r}$, which is a measure of the rate at which a anisotropic system relaxes towards isotropy.  

The rotational diffusional coefficient for a fibril of length $L$ in a polydisperse system close to isotropy,  could be appoximated as~\cite{marrucci1983,marrucci1984,Brouzet_2018,Brouzet_2019} 

\begin{equation}
D_{r} (L)\approx \frac{\tilde{\beta} k_{B} T L_{*}^{4}}{\eta L^{7}}, \label{rotdiff}
\end{equation}
where $\tilde{\beta}$ is a numerical factor $\simeq$ 1000~\cite{chow1985,pecora1985,rogers2005}, the Boltzmann constant $k_B= 1.38 \times 10 ^{-23}$~J$~$K$^{-1}$, the temperature $T= 300$~K, the solvent viscosity $\eta_{s}= 1 $~mPa$~$s.  The entanglement length $L_{*}$  is defined as~\cite{marrucci1983,marrucci1984} 

\begin{equation}
L_{*} = {\left( \int_{0}^{+\infty} \tilde{c}(L)L dL\right)^{-1/2},}
\end{equation}

\noindent where $\tilde{c}$ is the concentration distribution dependent on fibril length $L$. The entanglement length $L_{*}$ indicates if the fibrils are in  a dilute (not entangled, $cL$$^{3}$ $\ll$ 1, $c$ being concentration of fibrils , $L$ $ < $ $L_{*}$) or semi-dilute regime (1~$\ll$~$cL$$^{3}$~$\ll$ $L/d$, entangled, $L$ $ >$ $L_{*}$). More details related to $L_{*} $ can be found in Refs.~\cite{Brouzet_2018,Brouzet_2019}. For the present colloidal dispersion, $L_{*}$~$\approx$~45~nm, and all the fibril lengths $L$ $ >$ $L_{*}$, fall in the semi-dilute regime.  Substituting all the values in Eq.~(\ref{rotdiff}), for an average fibril length $L \sim 750 $ nm, the  rotational diffusional coefficient close to isotropy becomes $D_{r}$~$\approx$~0.126~rad$^{2}$~s$^{-1}$. Thus, $\dot{\gamma}_{crit}$ is approximated to be around $\dot{\gamma}_{crit}$ $=$ $6D_{r}$~$\simeq$~0.80 s$^{-1}$ as depicted in Fig.~\ref{fig:fig2}.

Thus, the low shear viscosity i.e. zero shear viscosity $\eta_0$ of the colloidal dispersion is considered near to the critical shear rate $\dot{\gamma}_{crit}$ estimated as per the Eq.~(\ref{rotdiff}) (see Fig.~\ref{fig:fig2}). The viscosity ratio between the core and sheath flow defined as $\chi = \eta_{1}/\eta_{2}~=~4500$  is arrived based on the zero shear viscosity $\eta_0$ of the core fluid i.e.  $\eta_{1}$~=~$\eta_0=4500$~mPa$~$s, and $\eta_2=1$~mPa$~$s, being the viscosity of sheath fluid. \\

\begin{centering}
\subsubsection{Péclet number estimation}\label{peclet number}
\end{centering}

In order to verify whether a \emph{de-facto} interface persists in the experimental system, time scale analysis is carried out by estimating the Péclet number based on the characteristic length scale $h$ of  the channel system~\cite{atencia2004,cubaud2006}. The translational diffusion coefficient~$D$ for  Brownian fibrils of average length $L \sim 750 $ nm  and diameter $d \sim 2.3$~nm diffused in a solvent of water is given as~\citep{doi1988}
\begin{equation}
D = \frac{k_{B} T \ln(L/d)}{2\pi \eta l},
\end{equation}
\noindent where the temperature $T= 300$~K, the Boltzmann constant $k_B= 1.38 \times 10 ^{-23}$~J$~$K$^{-1}$, and the solvent viscosity $\eta=~1~$mPa$~$s.  Substituting all the values,  $D$ becomes approximately $ 5 \times 10^{-12}$~m$^{2}~$s$^{-1}$. The Péclet number  is estimated as~ $Pe~=~Uh/D$~$~\approx~2~\times~10^{6}$ for an average flow velocity of $U \approx 10 $~mm$~$s$^{-1}$.  Thus, as the Péclet number is very large,  the time scale for the interdiffusion between the colloidal dispersion and its solvent is almost negligible compared to the convection time scale of the two fluids in the channel. Therefore, in such an event, a sharp \emph{de-facto} interface between the two fluids is likely to exist in the experiments~\cite{joseph1993fundamentals, atencia2004}. In the context of  the present experimental fluids, at  such a \emph{de-facto} interface, EIT is expected to be present~\cite{truzzolillo2015,truzzolillo2016}.

In addition, the rheological behaviour of the colloidal dispersion is also controlled by the fact that the nanofibrils form a percolating volume spanning arrested state at very low concentrations~\cite{rosen2020}.  This, in turn,  makes the diffusion of  nanofibrils into the surrounding sheath flows very slow compared to the time scales of the dynamics in the flow-focusing channel, making the Korteweg stresses long-lived. \\

\begin{centering}
	\subsubsection{Data acquisition method}\label{acquisition}
\end{centering}

 Three-dimensional core fluid thread topologies and the velocity field measurements are carried out by employing an light-based spectral domain optical coherence tomography. Optical coherence tomography (OCT) is a non-invasive volumetric imaging technique that uses a  broadband light source, and operates based on the principle of low-coherence interferometry~\cite{huang1991optical}. Utilising the Doppler principle, OCT can simultaneously capture the structural properties as well as the motion of opaque and turbid media sample with micron-level resolution~\cite{drexler2008optical,leitgeb2014doppler}. The  wavelength of  the light source of  the spectral domain OCT  used in the present work  is $1310$~nm with a bandwidth of $270$~nm, and a resolution of $\sim$3~$\mu$m in both the axial and transverse directions. More details related to the working principle of OCT, subsequent data acquisition and processing employed for the present study is described in Ref.~\citep{gowda2019effective}.

 In the present work, the 3-D experimental measurements are performed for all the flow-focusing configurations illustrated in Fig.~\ref{fig:fig1} with the above mentioned fluid properties. These experimental measurements will be used in Secs.~\ref{thread shapes} and \ref{centreline velocities} for cross-comparison with the numerical observations, and in Sec.~\ref{model curve} to measure the experimentally acting $\Gamma_e$ between the colloidal dispersion-solvent system. \\
 
\begin{centering}
{\subsection{Numerical setup} \label{Numerical set-up and methodology}}
\end{centering}

The numerical computations have been performed by utilising a recently developed finite volume based geometric volume of  fluid (VoF) method, an interface advection algorithm called isoAdvector~\cite{hirt1981volume, roenby2016computational}. The implementation of the  algorithm is incorporated in \textit{interIsoFoam}, a two-phase incompressible \textit{immiscible} open-source flow solver which is a part of the OpenFOAM\textsuperscript \textregistered community~\cite{OpenCFD}. The algorithm accurately captures and advects the sharp interface, a key aspect in the numerical computation of multiphase flows. The rationale behind choosing the \textit{immiscible} fluid solver was elucidated in the introduction, and will be again  clarified in the upcoming Sec.~\ref{model curve}. The set of equations being solved for an \textit{immiscible} system of two fluids are,

\noindent the continuity equation
\begin{equation}
\mbox{\boldmath $\nabla$}\cdot \,
\mbox{\boldmath $U$} = 0,
\label{continuity}
\end{equation} 

\noindent the Navier-Stokes equation together with the continuum representation of an interfacial tension force ${\mbox{\boldmath $F_s$}}$~\cite{Brackbill1992}
\begin{equation}
\frac{\partial
	{\rho_{b} \mbox{\boldmath $U$}}}{\partial t} +  \mbox{\boldmath $\nabla$}\cdot (\rho_{b}\mbox{\boldmath $U$}{\mbox{\boldmath $U$}} )= -{\mbox{\boldmath $\nabla$}} {p} + {\mbox{\boldmath $\nabla$}} \cdot {\mbox{\boldmath $T$}}+ {\mbox{\boldmath $F_s$}},
\label{momentum}
\end{equation}

\begin{equation}
{\mbox{\boldmath $F_s$}=\gamma \kappa(\mbox{\boldmath $\nabla$} \alpha)}, 
\end{equation}
where $\gamma$ is the interfacial tension and $\kappa$ is the curvature of the interface

\begin{equation}
{\kappa = - \mbox{\boldmath $\nabla$} \cdot \bigg(\frac{\mbox{\boldmath $\nabla$} \alpha}{|\mbox{\boldmath $\nabla$} \alpha|}\bigg)},
\end{equation}

\noindent and the equation for the advection of phase/volume fraction $\alpha$
\begin{equation}
\frac{\partial \alpha}{\partial t} + \mbox{\boldmath $\nabla$}\cdot (\alpha\mbox{\boldmath $U$})  = 0.
\label{diffDim}
\end{equation}

\noindent Here  $\textbf{\textit{T}}$ represents the deviatoric stress tensor, $\textbf{\textit{U}}$ is the velocity vector field and $p$ is the pressure field. The density $\rho_{b}$  and viscosity $\mu_{b}$ are computed as $\rho_{b} = \rho_{1}\alpha + {\rho_{2}} (1 -\alpha)$, $\mu_{b} = \mu_{1}\alpha + {\mu_{2}} (1 -\alpha)$
based on the weighted average distribution of the volume fraction $\alpha$ of  fluid where $\rho_1$, $\rho_2$, $\mu_{1}$, ${\mu_{2}}$ are the densities and the viscosities of the two fluids, respectively. 

In the present study, 3-D numerical computations are  performed for the geometrical configurations illustrated in Fig.~\ref{fig:fig1}.  At the channel walls, a no-slip velocity boundary condition and a contact angle of $\theta = 0^{\circ}$ is imposed for the phase field. A uniform velocity flow profile is prescribed at the core and sheath flow channel inlets  based on the flowrates of $Q_{1}$~=~6.5  mm$^{3}~$s$^{-1}$ and $Q_{2}$~=~ $7.5$~mm$^{3}~$s$^{-1}$. At the channel outlet, pressure is set to atmospheric, and zero gradient for the volume fraction. The non-Newtonian Carreau model depicted in Fig.~\ref{fig:fig2} is implemented for the rheology of the core fluid while the viscosity of water $\eta_2=1$~mPa$~$s is set for the sheath fluid.

The numerical computations are performed on Cartesian meshes. The size of the computational domain comprises of three inlet channels of length~$5h$ and an outlet channel of length $30h$  with a square cross-section of width $h$ = 1 mm.  The inlet channel domains are discretized with a unidirectional grid along the channel length and an equidistant grid spacing across the cross-section.  The outlet channel domain has an equidistant grid spacing of  $\Delta$ = 2$\times$$10^{-5}$ m ($\Delta x  = \Delta y = \Delta z$) both along the channel length and across the square cross-section.  An adaptive time step method is used to achieve the stability and convergence of the computations. The computations were run on $256$~processors, and the simulations took  $48$ h.  For  additional details concerning to numerics and grid convergence, we guide the interested reader to look upon Ref.~\cite{gowda2019effective}.

In application to two-phase microfluidic flows via hydrodynamic focusing, the solver has been thoroughly tested and validated through an extensive comparison with the experimental data. Excellent qualitative and quantitative agreement between numerical and experimental results~\cite{gowda2019effective} is demonstrated, along  with the capture of diverse flow regimes observed by Ref.~\cite{Cubaud_2008} in a typical microfluidic flow-focusing setup. \\

\begin{centering}
\section{Results and discussion}\label{Results}
\end{centering}

We first carry out a detailed numerical and experimental investigation of the effect of confluence angle $\beta$ on the 3-D thread morphology and shape of the wetted region. Then, in Sec.~\ref{model curve},  experimentally measured thread height~$\varepsilon_z/h$ as in the case of reference flow-focusing configuration ($\beta = 90\degree$) is compared with the master curve to estimate the  $\Gamma_{e} $  acting between the present experimental fluids.

The flow rates and rheologies of fluids in the computations and experiments are set as given earlier in Secs.~\ref{geom-setup} and \ref{fluids}. \\

\begin{centering}
\subsection{Morphology of the thread and shape of the wetting region}\label{thread shapes}
\end{centering}

In this section, we compare the 3-D experimental thread shape measured with OCT for geometrically varying flow-focusing configurations described in Sec.~\ref{exp-setup}, and those obtained by numerical computations for the corresponding configurations. All the numerical computations are performed with the EIT, $\Gamma_{e}=\gamma= 0.615$~mN$~$m$^{-1}$. This choice will be motivated in the upcoming  Sec.~\ref{model curve}.

\begin{figure} [tbp!]
	\centering
	\includegraphics[width=1\textwidth]{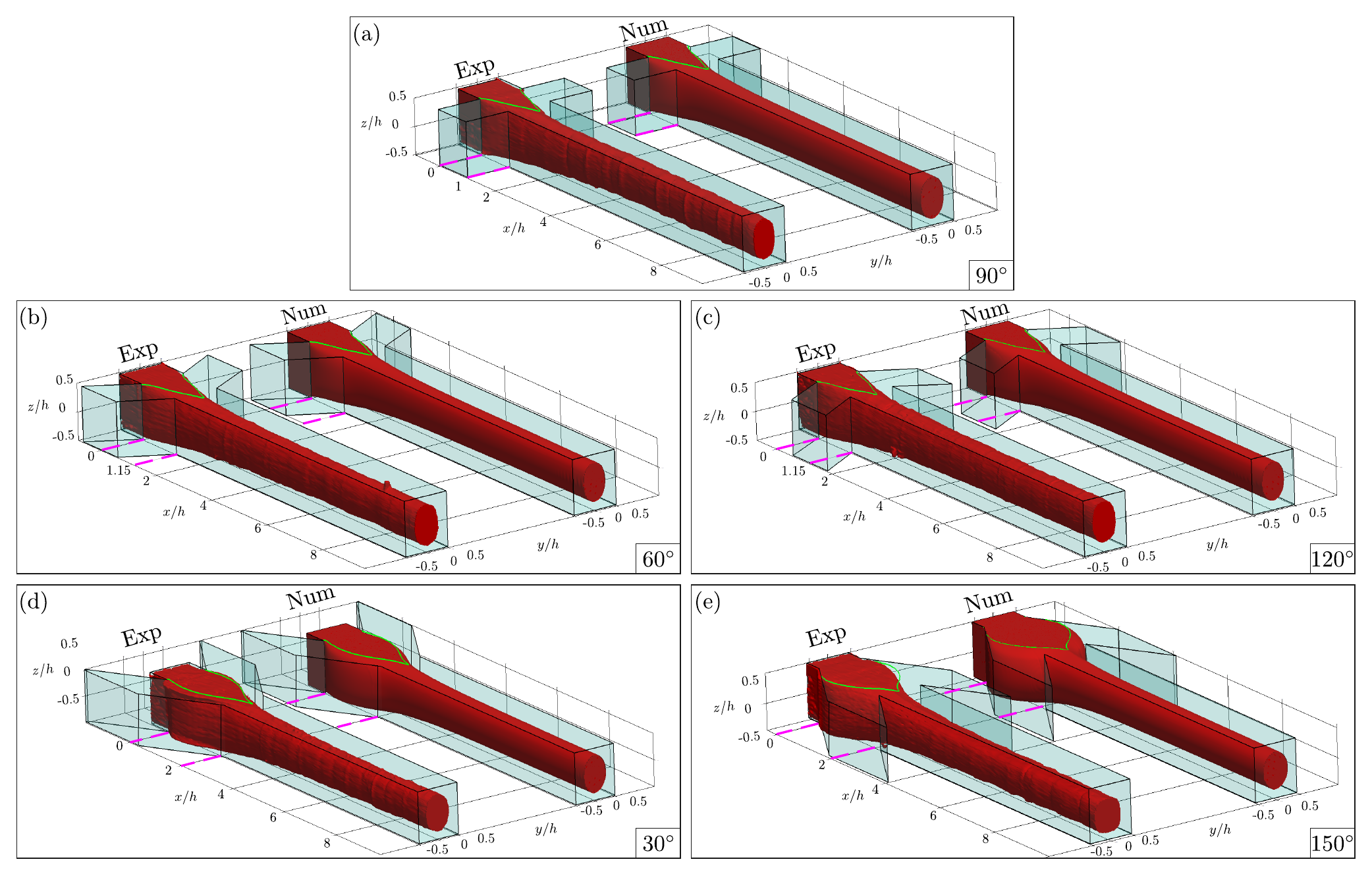}
	\caption{(a) - (e) 3-D view of the experimental and numerical threads for flow-focusing configurations with varying confluence angle $\beta$. The green curves at the top planes $z/h=0.5$ on all the five panels represent the boundary of  wetted region created by the core fluid before the detachment. The horizontal dashed magenta lines in all the five panels indicate the confluence region for the respective configurations.}   
	\label{fig:fig4_Octthread}
\end{figure}

\begin{figure} [tbp!]
	\centering
	\includegraphics[width=1\textwidth]{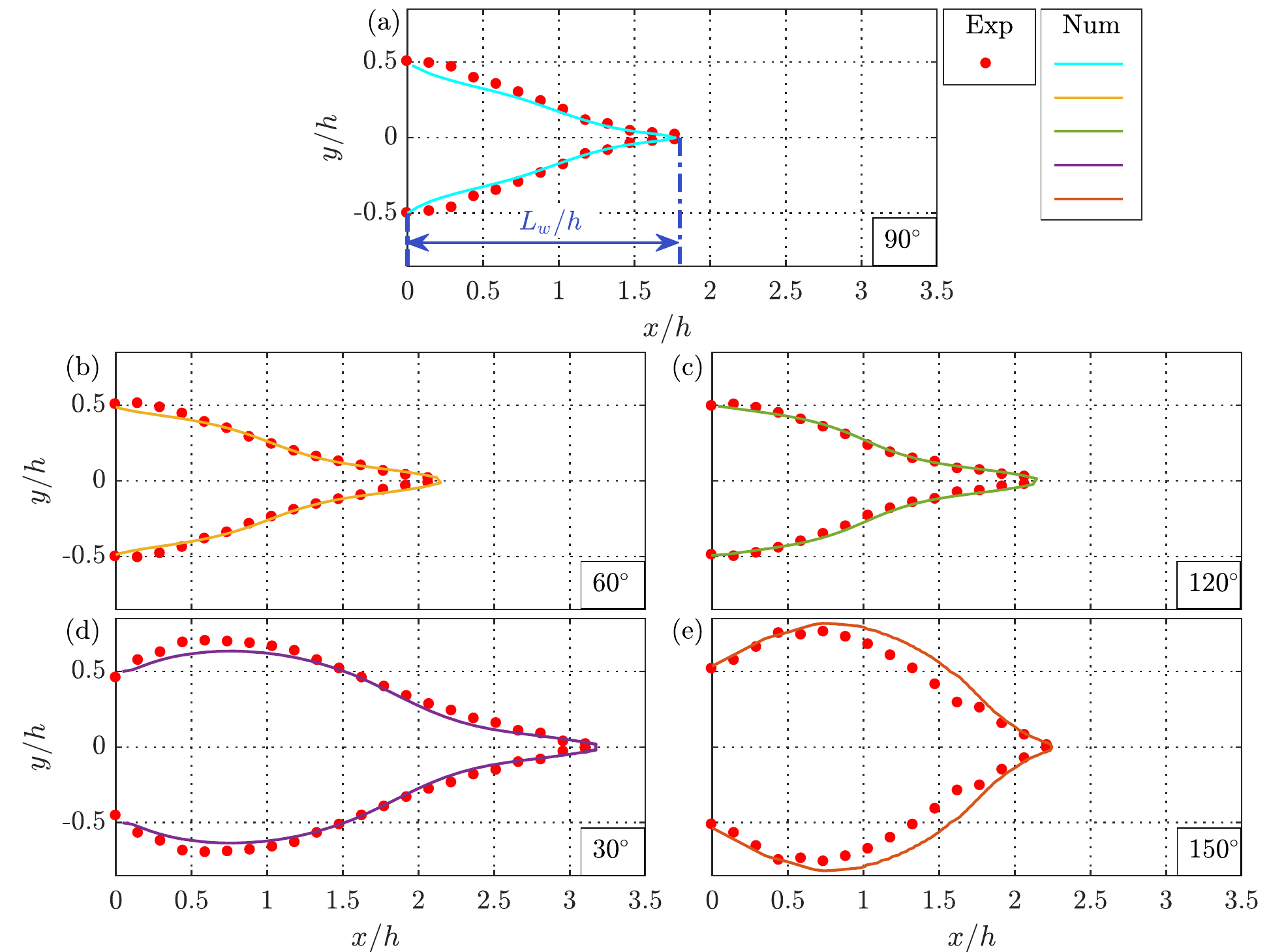} 
	 \caption{(a) - (e) Top view of the experimental and numerical wetted region 
		morphologies of the core fluid in the plane $z/h=0.5$ for flow-focusing configurations with varying confluence angle~$\beta$.  The distance from $x/h=0$ to the point of detachment of core fluid from the top wall is referred to as `wetted length'~$L_w/h$, indicated in panel~(a). The wetted region shape and length~$L_w/h$ vary as per the respective geometrical configuration. These wetted boundaries  are also shown in the previous Figs.~\ref{fig:fig4_Octthread}(a) - \ref{fig:fig4_Octthread}(e)  at the top plane $z/h=0.5$ marked with green curves. }
	\label{fig:fig5_Wetting_length}
\end{figure}

Firstly, a few generic features applicable to all the geometrical configurations are noted before moving into a geometry specific detailed discussion. 
Figures~\ref{fig:fig4_Octthread}(a) - \ref{fig:fig4_Octthread}(e) show the experimental and numerical 3-D thread shapes of the colloidal dispersion for various confluence angles~$\beta$. The qualitative agreement between experiments and numerics is ascertained to be quite good. A closer view shows  the core and sheath flow channel walls intersect at the point of confluence corresponding to $x/h=0$.  The region beginning from $x/h=0$ to the position where the confluence of core and sheath flow channels end (indicated by dashed magenta lines) is referred to as the \emph{confluence region}. However, the length of  confluence region varies depending on the confluence angle~$\beta$.  The confluence region is shortest (0~$\leq$~$x/h$~$\leq$~1) for the reference configuration ($\beta$~=~90$\degree$) and longest (0~$\leq$~$x/h$~$\leq$~2) for  $\beta=[30$\degree$, 150$\degree$]$ pair. 
Meanwhile, at the top and bottom walls of the channel (i.e. at $z/h =\pm 0.5$), the colloidal dispersion invariably remains attached to the walls even after $x/h$ $>$ 0. The dispersion continues to stay attached to the upper and lower walls upto a pinch-off point. The region originating from $x/h$ $=$ 0 to the colloidal thread detachment point is denoted the \emph{wetted region} and the subsequent length is called \emph{wetted length $L_w/h$} (see Fig.~\ref{fig:fig5_Wetting_length}(a)). Indeed, both the  wetted region and the wetted length vary with $\beta$  as observed in Figs.~\ref{fig:fig4_Octthread}(a)~-~\ref{fig:fig4_Octthread}(e) (marked by green curves at the top planes $z/h=0.5$),  and more clearly in Figs.~\ref{fig:fig5_Wetting_length}(a)~-~\ref{fig:fig5_Wetting_length}(e). Also here, a very good agreement is notable between the computations and experiments.

\begin{figure} [tbp!]
	\centering
	\includegraphics[width=1\textwidth]{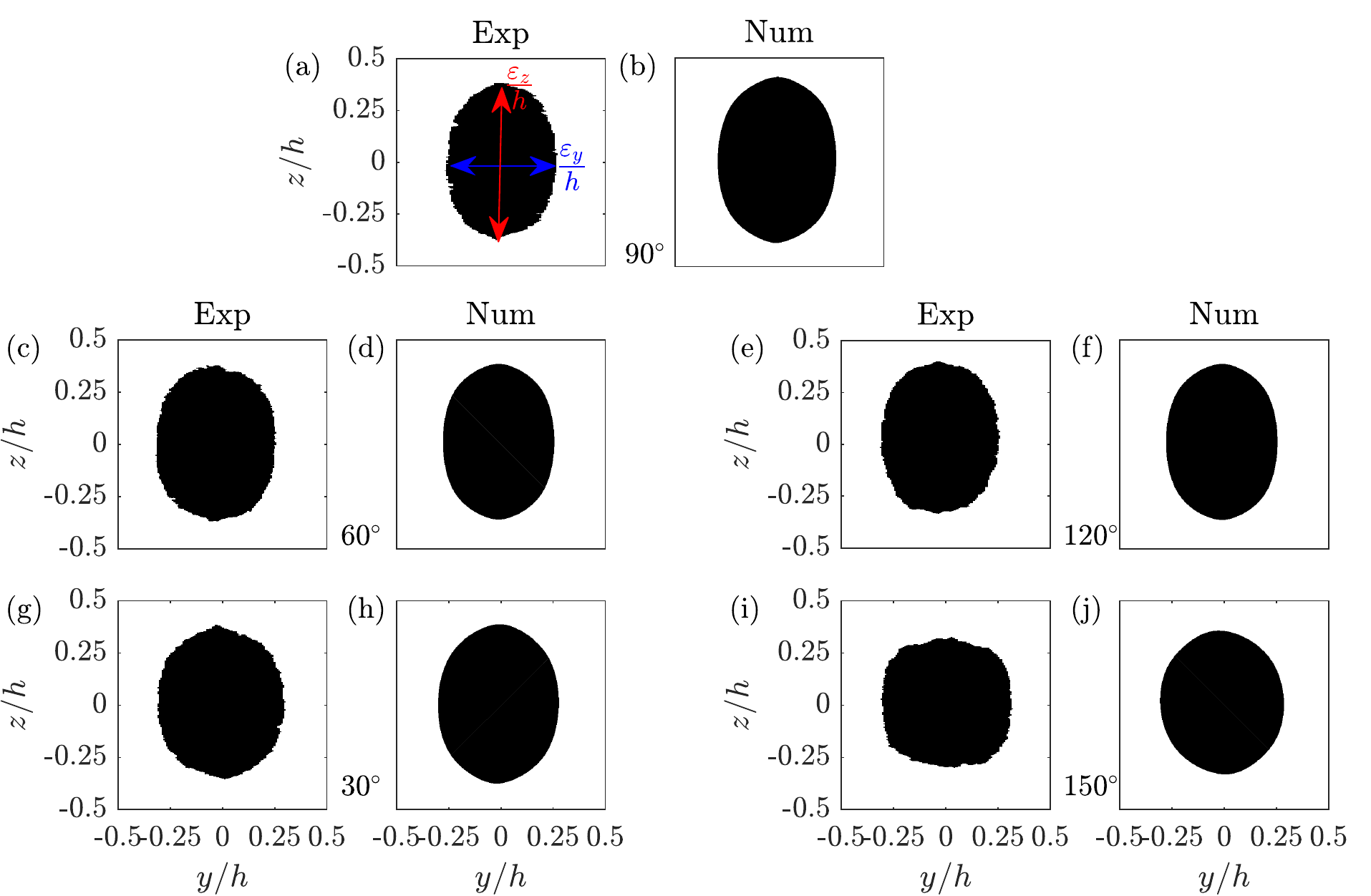} 
\caption{Experimental (panels (a),(c),(e),(g),(i)) and numerical (panels (b),(d),(f),(h),(j)) thread cross-sections at downstream position $x/h = 10$ for flow-focusing configurations with varying confluence angle~$\beta$. Black color denotes the core fluid while white color represents the sheath fluid.  Thread width~$\varepsilon_y/h$ (blue) and height~$\varepsilon_z/h$ (red) denoting the ellipsoid axes is pictured in panel (a).}\label{fig6:Cross-section}  
\end{figure}

\begin{figure} [tbp!]
	\centering
	\includegraphics[width=1\textwidth]{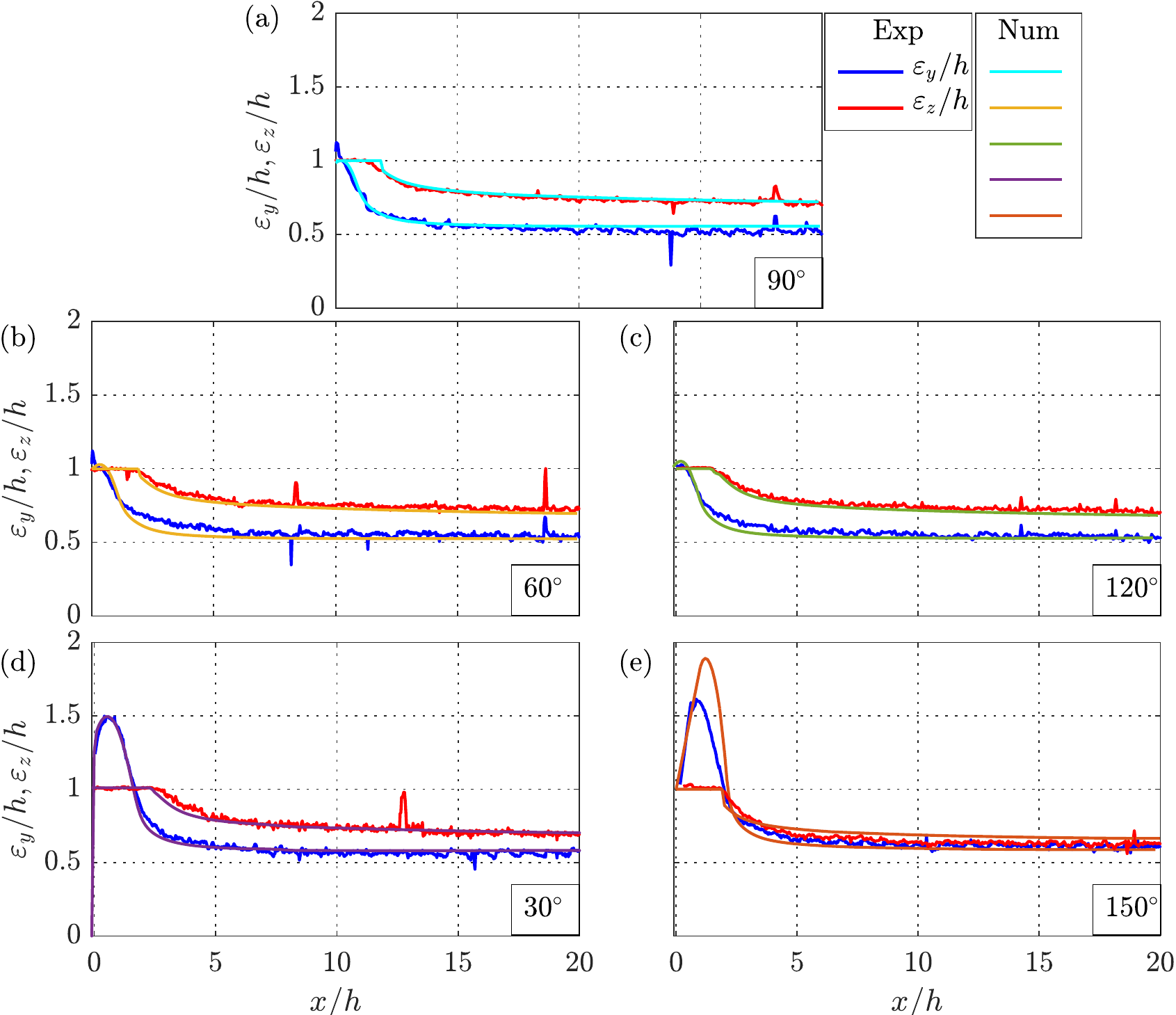} 
	\caption{(a) - (e) Thread width $\varepsilon_y/h$ and height $\varepsilon_z/h$  as a function of downstream positions $x/h$ for flow-focusing configurations with varying confluence angle $\beta$. The experimental data are represented in blue and red color, while the numerical data are denoted by various solid color lines shown in the rightmost legend in panel~(a) as per the respective geometrical configuration.}
	\label{fig:fig7_Threadwidth_height}
\end{figure}

Further, as observed from Figs.~\ref{fig:fig4_Octthread}(a) - \ref{fig:fig4_Octthread}(e), the evolution of streamwise thread development appears to be affected by $\beta$ in the upstream while at far downstream the thread shapes seem to be nearly elliptical.  The cross-sectional views of the elliptical thread at downstream position $x/h=10$ are displayed in Figs.~\ref{fig6:Cross-section}(a) - \ref{fig6:Cross-section}(e) with its major and minor axes denoted by $\varepsilon_z/h$, $\varepsilon_y/h$. Besides, quantitatively,  the excellent agreement is exemplified between experimentally measured height~$\varepsilon_z/h$ (red curves) and width~$\varepsilon_y/h$ of the thread (blue curves) with the numerically computed thread height and width (denoted by various color curves as per the respective configuration) depicted in Figs.~\ref{fig:fig7_Threadwidth_height}(a)~-~\ref{fig:fig7_Threadwidth_height}(e).

In view of the above observations, it is apparent that the confluence angle $\beta$ has a significant influence both on the wetted region morphology and the thread development. Indeed, characterising the influence of $\beta$ is an important aspect from the hydrodynamic assembly of nanofibrils point of view~\citep{haakansson2014hydrodynamic}, since the shape of thread regulates the cross-section of an assembled material. This, in turn, could enhance the scope for synthesizing materials of complex shapes such as rods, ellipsoids, discs and so on. Hence, in what follows, is a systematic analysis of the effect of confluence angle $\beta$.

The flow rates, fluids rheology and channel cross-sectional width~\textit{h} across all the flow-focusing configurations are held fixed. So, the parameter that  varies due to the effect of $\beta$, is the sheath flow momentum in the parallel and normal directions to the core flow at the confluence region. Therefore, a qualitative understanding can be gained by centering on the sheath fluid momentum and in turn, its effect on the core fluid thread topologies.

Beginning with the reference configuration ($\beta$~=~90$\degree$), the sheath flows are perpendicular to the core flow. Accordingly, sheath flow momentum is acting only normal to the core flow over the confluence region 0~$\leq$~$x/h$~$\leq$~1, and the sheath fluid impinges the core fluid with maximum impact in the normal direction as seen in Fig.~\ref{fig:fig4_Octthread}(a). As a result, the colloidal thread detaches  with a shorter wetted length $L_w/h\approx1.8$ (Fig.~\ref{fig:fig5_Wetting_length}(a)) compared to other configurations (Figs.~\ref{fig:fig5_Wetting_length}(b) - \ref{fig:fig5_Wetting_length}(e)). 
In fact, as it can be seen from Fig.~\ref{fig:fig7_Threadwidth_height}(a), even the thread width~$\varepsilon_y/h$ decreases much faster than the height~$\varepsilon_z/h$ up to $x/h\approx6$, highlighting the extent of impact of sheath flow momentum along the streamwise thread development.  Far downstream the width attains a constant value. However,  after the thread detachment, the effect of sheath flow momentum on the height is less significant. Furthermore, since the sheath flow momentum is maximized in the direction normal to the core flow  when $\beta=90\degree$, all the upcoming geometrical configuration discussions will be relative to this reference configuration. In particular when the sheath flow momentum normal to the core flow is taken into consideration, and hence the choice of  geometrical configurations  placement in the order as shown in Fig.~\ref{fig:fig1} were opted.

Continuing  to  the $\beta=[60$\degree$, 120$\degree$]$ pair of configurations, the sheath flow momentum acts both in parallel and normal to the streamwise core flow at their respective confluence regions as seen in Figs.~\ref{fig:fig4_Octthread}(b) and \ref{fig:fig4_Octthread}(c). Most importantly, both the configurations have the same length of  confluence regions i.e. 0~$\leq$~$x/h$~$\leq$~1.15. However, in the $\beta=60\degree$ case, the sheath flow is impinging the core flow in the same direction as the streamwise core flow, which can be refered to as positive impingement, while in the $\beta=120\degree$ case, the sheath flow is impinging opposite to the direction of the streamwise core flow, i.e.~a negative impingement.  For the purpose of clarity, top views of  both the configurations is illustrated in Figs.~\ref{fig:fig1}(b) and \ref{fig:fig1}(c), which gives a better notion of positive and negative impingement (following the signs of arrow representing the core and sheath flows). Surprisingly, the thread morphology and shape of the wetted region for these two cases closely resembles each other as seen in Figs.~\ref{fig:fig4_Octthread}(b), \ref{fig:fig5_Wetting_length}(b) and \ref{fig:fig4_Octthread}(c), \ref{fig:fig5_Wetting_length}(c) in spite of difference in the fundamental impingement direction. As it can be observed from Figs.~\ref{fig:fig5_Wetting_length}(b) and \ref{fig:fig5_Wetting_length}(c), in both the cases, the wetted area increases and the wetted length extends up to $L_w/h\approx2.1$, which is much longer than the reference configuration $L_w/h$  (Fig.~\ref{fig:fig5_Wetting_length}(a), $L_w/h\approx 1.8$). This indicates that the main effect of $\beta$ is through the sheath flow momentum normal to the core flow, which has been weakened by about 13.4\% in relative to the reference configuration in both cases. Further, during the development of the thread width~$\varepsilon_y/h$  and height~$\varepsilon_z/h$ as a function of downstream positions~$x/h$, both the configurations exhibit similar characteristics as displayed in Figs.~\ref{fig:fig7_Threadwidth_height}(b) and \ref{fig:fig7_Threadwidth_height}(c).  In both cases, the width and height of the thread decay slightly faster up to $x/h\approx6$  and thereafter, stays almost  constant far downstream with an elliptical cross-section as visualized in Figs.~\ref{fig6:Cross-section}~(panels (c),(d) and panels (e),(f)).

On the other hand, for  the $\beta=[30$\degree$, 150$\degree$]$ pair, the thread shape and wetted region morphologies differ substantially as viewed in Figs.~\ref{fig:fig4_Octthread}(d),~\ref{fig:fig5_Wetting_length}(d) and \ref{fig:fig4_Octthread}(e),~\ref{fig:fig5_Wetting_length}(e).   Figures~\ref{fig:fig4_Octthread}(d) and \ref{fig:fig4_Octthread}(e) show the length of confluence regions 0~$\leq$~$x/h$~$\leq$~2 being larger than in other configurations  (Figs.~\ref{fig:fig4_Octthread}(a) or \ref{fig:fig4_Octthread}(b),~\ref{fig:fig4_Octthread}(c)). Notably, in both cases, there is an expansion of the colloidal thread symmetrically in the transverse or sheath flow direction ($y$-direction) near the confluence region. This could be due to the decrease in the impact of sheath flow momentum normal to the core flow. Compared to the reference configuration, the magnitude of  normal sheath flow momentum  is reduced by around 50\% due to the effect of $\beta$. In the case of postive impingement of sheath flow ($\beta~$=~30$\degree$) there is a small expansion of the wetted area first and the wetted length extends up to $L_w/h\approx3.1$ as depicted in Fig.~\ref{fig:fig5_Wetting_length}(d). However, for  the negative impingement ($\beta$~=~150$\degree$),  the wetted area expands considerably by curving outward in the transverse direction with a much shorter wetted length $L_w/h\approx2.1$ as seen from Fig.~\ref{fig:fig5_Wetting_length}(e).

The evolution of streamwise thread development in Figs.~\ref{fig:fig7_Threadwidth_height}(d) and \ref{fig:fig7_Threadwidth_height}(e) shows that the width~$\varepsilon_y/h$ of the thread in the confluence region  0~$\leq$~$x/h$~$\leq$~2 overshoots up to $\varepsilon_y/h\approx1.5$ ($\beta$ =~30$\degree$ case) and even more for the negative impingement ($\beta$ =~150$\degree$) before the decay, highlighting the expansion of the thread in the tranverse direction.  Another striking behaviour is for the $\beta$ = 150$\degree$ configuration, where the experimental cross-section of the thread is somewhat different from the elliptical shape as illustrated in Figs.~\ref{fig6:Cross-section}(i), (j). In fact, the numerical curve shows an overprediction in the upstream near the confluence junction during the expansion. This  feature of experimental thread cross-section in Fig.~\ref{fig6:Cross-section}(i)  is not accuratley captured by the numerics. \\

\begin{centering}
	\subsection{Estimation of  effective interfacial tension}\label{model curve}
\end{centering}

In this section, we first present an overview of the model used to obtain  a master curve that was proposed at the conclusion of our previous study in Ref~\cite{gowda2019effective}. Then, we fit the master curve to an exponential function, and employ it here to estimate the  $\Gamma_e$ between the present experimental fluids.

\begin{figure} [tbp!]
	\centering
	\includegraphics[width=1\textwidth]{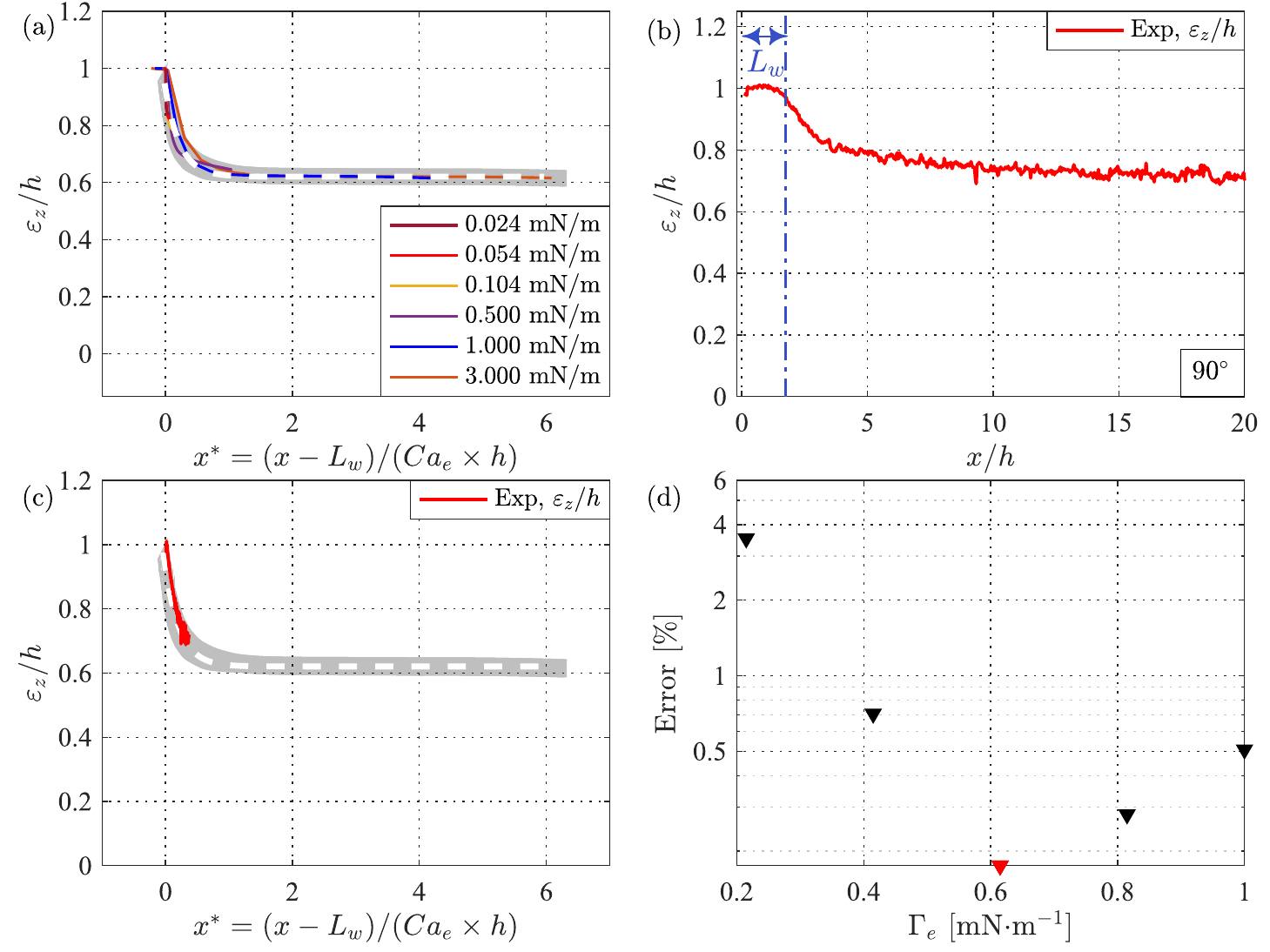}
	\caption{ (a): The numerical thread heights~${\varepsilon_z/h}$ at  various interfacial tensions~$\gamma$ plotted as a function of scaled downstream length $x^{*}$ (detailed in text). The data collapse on to a master curve.  These simulated data are taken from  Ref~\cite{gowda2019effective}. The master curve can be best represented by an exponential fit [see Eq.~(\ref{fit})] denoted by the dashed-white line.   The fit parameters $a$, $b$ and $c$ are reported in Table~\ref{tab1}.  Panel~(b) shows the experimental thread height~${\varepsilon_z/h}$ as a function of downstream positions $x/h$ measured in the reference configuration ($\beta$~=~90$\degree$) of the present study. The dash-dotted vertical blue line in panel~(b) indicates the measured wetted length of the thread before the detachment ($L_w/h$~$\sim$~1.8). In panel~(c), the experimental thread height~${\varepsilon_z/h}$ is compared with the exponential fit (dashed-white line) representing the master curve to estimate the $\Gamma_e$ acting between the present experimental fluids. The light gray backdrop in panels (a) and (c) is drawn for the purpose of  better visualization of  the plotted data. Panel~(d) shows the percentage variation of the error (defined in text) between the numerical and experimental thread heights for a range of  $\Gamma_e$ values near to the estimated $\Gamma_e$ (Table~\ref{tab1}) obtained from the master curve fit in panel (c).  The red marker in panel~(d) indicate the  $\Gamma_e$ that gives the best match with the experimental measurement. }   
	\label{fig:fig3_fittingcurve}
\end{figure}

Figure~\ref{fig:fig3_fittingcurve}(a) summarizes our observations performed through an extensive set of numerical computations in Ref~\cite{gowda2019effective}. These computations were  performed with the reference flow-focusing setup for a set flow rate and the given rheology of the fluids~\cite{gowda2019effective}.  As mentioned in the introduction, employing an \textit{immiscible} fluid solver, we had demonstrated that soon after the core fluid detachment from the top and bottom channel walls of the channel at $x$ = $L_{w}$,  the thread height varied from non-circular  (near-ellipsoidal) cross-section to a circular shape over a range of interfacial tensions $0.024$~mN$~$m$^{-1}< ~\gamma< ~3.000$~mN$~$m$^{-1}$. The larger the value of ~$\gamma$,  the faster the thread converges towards a circular shape for which the thread height attains a constant value of ~${\varepsilon_z/h}$ $\sim$ $0.63$.  

Following~\cite{gowda2019effective}, using appropriate notations, the typical time and length scales for a near-ellipsoidal thread to approach circular shape can be written~as,

\begin{equation}
\tau=\frac{\eta_1}{\delta P}\propto\frac{\eta_1}{\gamma},\label{eq:model1}
\end{equation}

\begin{equation} 
l_r=U \tau \propto \frac{\eta_1 Q_1}{\gamma h^2}. \label{eq:model2}
\end{equation}

\noindent where $U$, $Q_1$ and $\eta_1$  are the velocity, flow rate and dynamic viscosity  of the core fluid. $\delta P$ is the pressure gradient dependent on the thread geometry  and is proportional to the interfacial tension $\gamma$. Comparing Eq.~(\ref{effective capillary}) and Eq.~(\ref{eq:model2}),  $\eta_1 Q_1/(\gamma h^{2})$ $\simeq$ $\eta Q/(\Gamma_e h^2)$ = $Ca_e $, the 
effective capillary number  where the modelling interfacial tension $\gamma$~$\equiv$~$\Gamma_e$ (EIT) in experiments.

Thus, from Eqs.~(\ref{eq:model1}) and (\ref{eq:model2}),  a scaled downstream length $x^{*}$ can be obtained by renormalizing the downstream length $(x-L_w)$ with $l_r$ leading  to $x^{*}$ = $(x-L_w)/(Ca_e~\times~h)$, respectively. As depicted in Fig~\ref{fig:fig3_fittingcurve}(a),  all the simulated thread heights (${\varepsilon_z/h}$) at various interfacial tensions $\gamma$  collapse  well on a master curve when plotted with the scaled co-ordinate $x^{*}$.

The master curve depicted in Fig.~\ref{fig:fig3_fittingcurve}(a) can be best fitted to an exponential function denoted by dashed-white line with the fitting parameters $a$, $b$ and $c$ given in Table~\ref{tab1}. 
 
 \begin{equation} \label{fit}
 {\varepsilon_z/h} = a\times e^{-bx^{*}} + c.
 \end{equation}
 
  The light gray backdrop in  Figs.~\ref{fig:fig3_fittingcurve}(a) and \ref{fig:fig3_fittingcurve}(c) is in place solely for the sake of  clarity in the graphics of the plots.  The fit parameters are evaluated through non-linear least-square regression performed in Matlab.
 
 To estimate the $\Gamma_e$ acting between the present  colloidal dispersion - solvent system, we use the spatial evolution of thread height $\varepsilon_z/h$ measured experimentally with the reference flow-focusing configuration ($\beta$~=~90$\degree$) as depicted in Fig.~\ref{fig:fig3_fittingcurve}(b), and compare with the exponential fit representing the master curve as shown in Fig.~\ref{fig:fig3_fittingcurve}(c).  By solving Eq.~(\ref{fit}) together with the fit parameters reported in Table~\ref{tab1}, the scaled downstream length $x^{*}$ is obtained.  As noted from above, the expression for $x^{*}$ is given as $x^{*}$ = $(x-L_w)/(Ca_e \times h)$. Substituting all the experimentally measured variables in $x^{*}$ and $Ca_e$ such as the core fluid flow rate  $Q_{1}$~=~6.5 ~mm$^{3}~$s$^{-1}$, viscosity of the core fluid, $\eta_1=4500$~mPa$~$s, the wetted length $L_w$~$\sim$~$1.8h$ along with the streamwise downstream positions $x$ of the thread height, and the channel width~$h$, the estimated $\Gamma_e \simeq 0.765$~mN$~$m$^{-1}$ can be retreived. 
 
\begin{table}[tbp!]
	\centering
	\caption{ Fitting parameters $a$, $b$  and $c$ of the exponential fit [Eq.~(\ref{fit}), dashed-white line in Figs.~\ref{fig:fig3_fittingcurve}(a) and \ref{fig:fig3_fittingcurve}(c)] representing the master curve.  The estimated $\Gamma_e$ acting between the present experimental fluids is obtained by utilising the spatially measured experimental thread height ${\varepsilon_z/h}$ ($\beta$~=~90$\degree$)  (shown in Fig.~\ref{fig:fig3_fittingcurve}(b)) and Eq.~(\ref{fit}) along with the fit parameters, and other  experimentally acquired quantities such as viscosity ($\eta_{1}$), flow rate ($Q_{1}$) of  the core fluid and the channel width $h$.}
	\label{tab1}
	\addtolength{\tabcolsep}{3pt} 
	\begin{tabular}{cccc}
		\hline
		\hline \\
		
		$a$ & $b$ & $c$ & Estimated $\Gamma_e$ (mN$~$m$^{-1}$)\\
		\tabularnewline
		\hline \\
		
		$~0.240~\pm~0.0008$  & $~2.950~\pm0.015~$  & $0.622~\pm0.0006~$ & $0.756~\pm0.0005~$\\
		
		\hline
		\hline
		
	\end{tabular}
\end{table}

Furthermore, as a verification, the numerical computations were performed in close proximity to the estimated $\Gamma_e$ utilising the reference flow-focusing configuration ($\beta$~=~90$\degree$).  The difference between the numerical and experimental~(Fig.~\ref{fig:fig3_fittingcurve}(b))  thread height results are plotted in Fig.~\ref{fig:fig3_fittingcurve}(d) as an error, which is defined as,
 
 \begin{equation}
 \delta_{\varepsilon_z}(\Gamma_e)=\sum_{i=1}^{N}{\left\lvert\frac{\varepsilon_{z,i}^{num}-\varepsilon_{z,i}^{exp}}{\varepsilon_{z,i}^{exp}}\right\lvert^{2},}
 \end{equation}

\noindent where $i = 1$ to $N$ are the downstream positions at which thread heights~$\varepsilon_z/h$ are evaluated after the thread detachment.  The minimum of  $\delta_{\varepsilon_z}(\Gamma_e)$ is indicated by a filled red symbol in Fig.~\ref{fig:fig3_fittingcurve}~(d) and occur at $\Gamma_e = 0.615$~mN$~$m$^{-1}$ which, in turn, affirms the estimated $\Gamma_e$ to be good with fairly accurate  order of magnitude. This value of  $\Gamma_e = \gamma$ is used in the numerical simulations of  all the geometrically varying flow-focusing setups as discussed in the previous Sec.~\ref{thread shapes}.  
 
 An interesting observation worth noting here is, that the value of  $\Gamma_e $ ($\mathcal{O}$$\sim$$10^{-1}$~mN$~$m$^{-1}$) for the present study colloidal dispersion-solvent system is almost one decade higher in magnitude than our previously evaluated value ($\mathcal{O}$$\sim$$10^{-2}$~mN$~$m$^{-1}$) for  a similar, but not identical  colloidal dispersion-solvent system~\cite{gowda2019effective} at the same concentration.  Both the colloidal dispersions exhibit non-Newtonian shear thinning behaviour, and the zero shear viscosity for the present case is  $\eta_{1}$~=~$\eta_0 \simeq 4500$~mPa$~$s while it was  $1750$~mPa$~$s  in Ref.~\cite{gowda2019effective}. 
 The order of magnitude difference in $\Gamma_e $ could be attributed to the variation in length fraction of nanofibrils, interfibril interactions leading to different rheological properties of the colloidal dispersion, in turn to that on variant dispersion-solvent  \emph{de-facto} interface properties. Indeed, this also concurs with the observations of  experimental studies performed by Refs.~\cite{truzzolillo2016, truzzolillo2015}, where $\Gamma_e $ between colloidal dispersions and its own solvent  were measured, and the variations in $\Gamma_e $ span over 5 decades for mild changes in the composition. \\

 \begin{centering}
\subsection{Centreline velocities} \label{centreline velocities}
\end{centering}

In addition to the evolution of thread shape, quantitative understanding of the flow field behaviour is vital for the modeling and prediction of hydrodynamic alignment of nanofibrils~\cite{jeffery1922}. Alignment of nanofibrils is a key factor in controlling and tuning the material properties~\cite{H_kansson_2016} of assembled materials.

\begin{figure} [tbp!]
	\centering
	\includegraphics[width=1\textwidth]{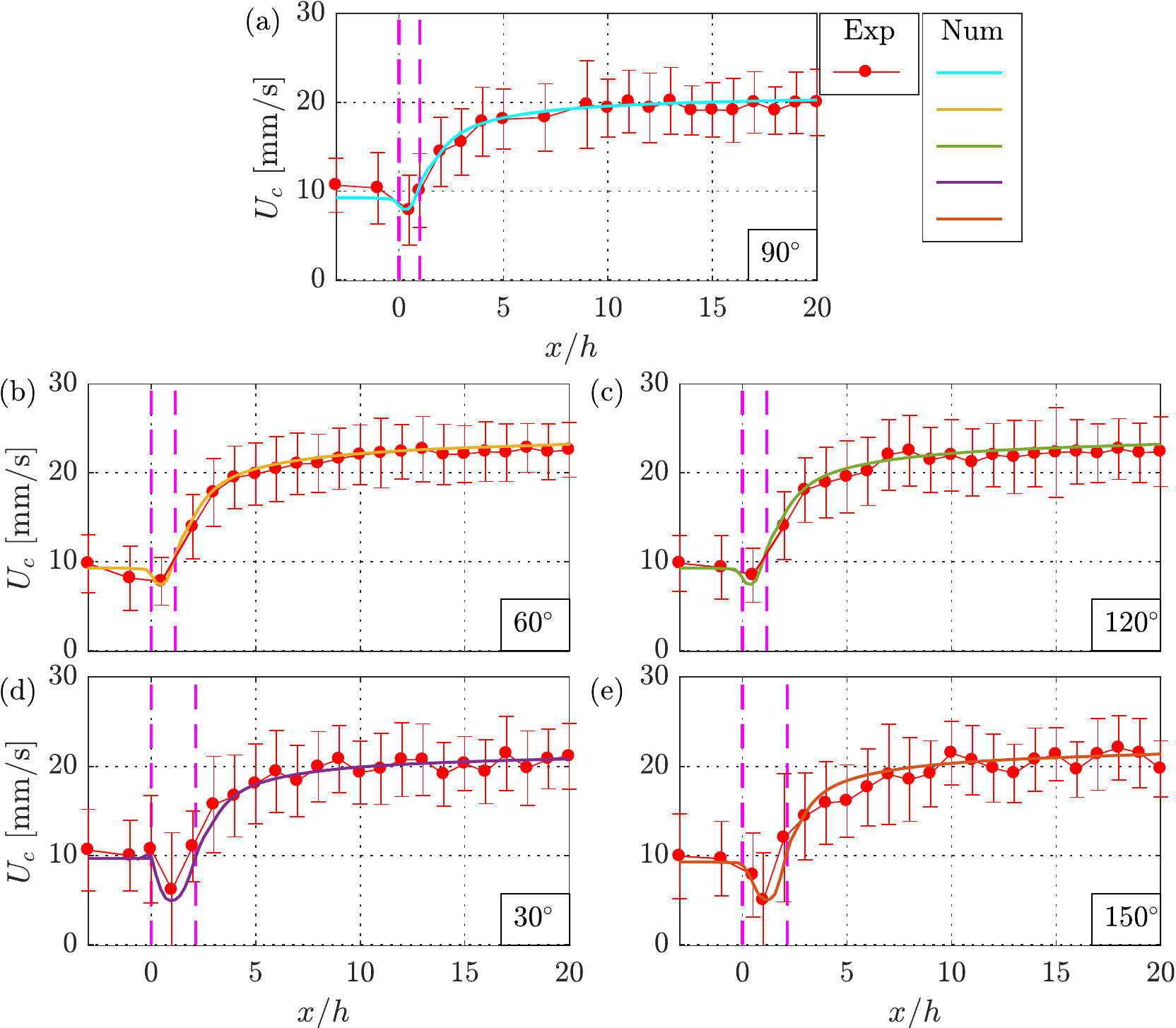}
\caption{Panels~(a) - (e) show the experimental and numerical centreline velocity~$U_c$  as a function of streamwise downstream positions~$x/h$ for flow-focusing configurations with varying confluence angle~$\beta$. The vertical dashed lines (magenta) on all the panels denote the confluence region of  the respective configurations. Error bars represent the standard deviation of experimental measurements at each position.}
	\label{fig:fig7_centrelinevelocity}
\end{figure}

\begin{figure} [tbp!]
	\centering
	\includegraphics[width=1\textwidth]{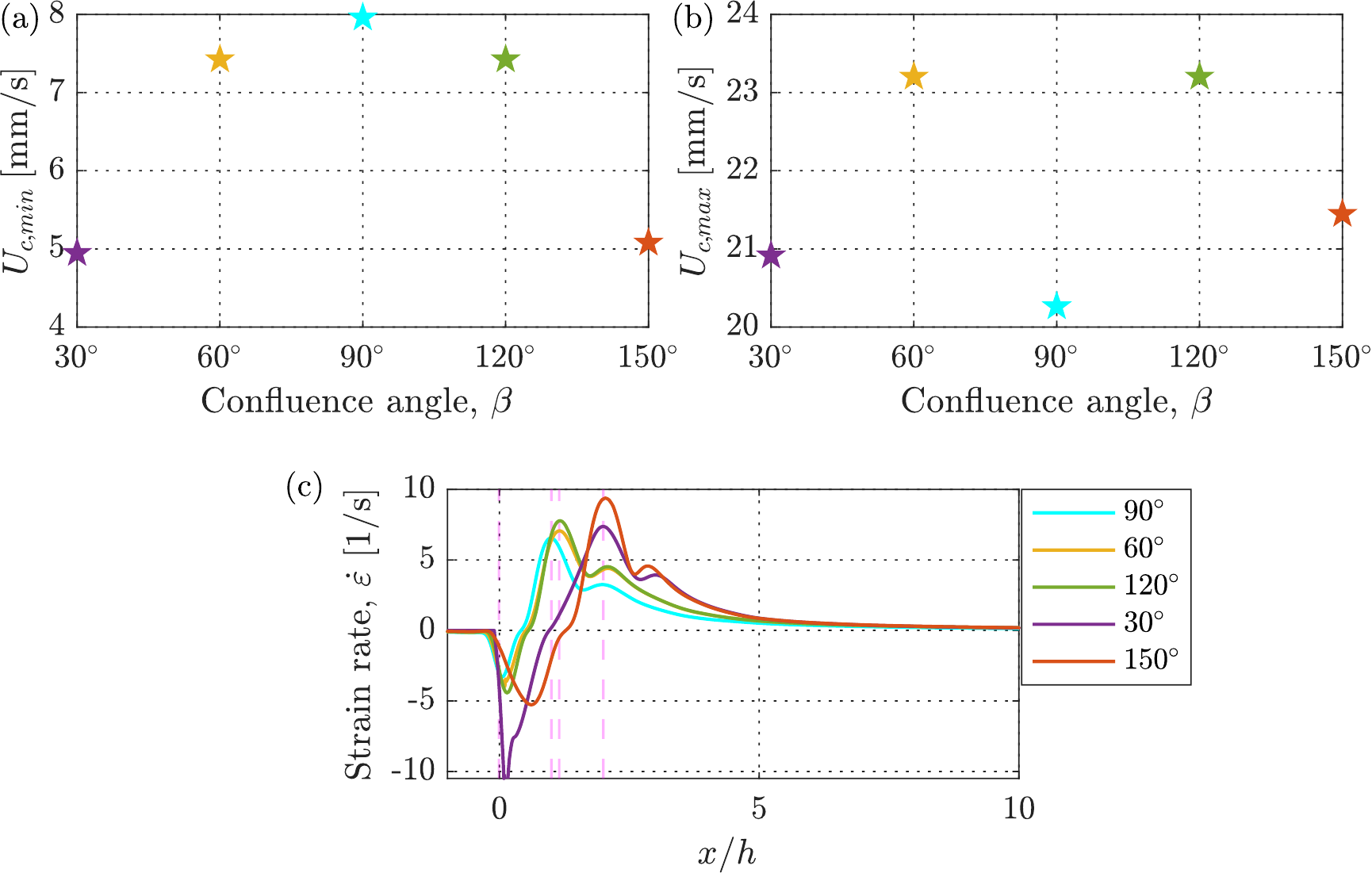}
	\caption{Numerical plots of  variation of minimum (panel (a)) and maximum velocity (panel (b)), and strain rate $\dot{\varepsilon}$ (panel (c))  along the centreline  as a  function of confluence angle~$\beta$.  The vertical dashed lines (magenta)  in panel~(c) denote the confluence regions  for all the configurations. The confluence region starts at $x/h$ = 0 for  all the configurations.}
	\label{fig:fig9_acceleration}
\end{figure}

Figure~\ref{fig:fig7_centrelinevelocity} shows the velocity variation along the centreline of the colloidal thread as a function of downstream positions~$x/h$ for all the five confluence angle geometries. The agreement between numerical computations and experimental measurements is excellent. Dashed vertical lines (magenta)  in all the panels (Figs.~\ref{fig:fig7_centrelinevelocity}(a) - ~\ref{fig:fig7_centrelinevelocity}(e)) mark the confluence regions. The trend of the centerline velocity ($U_c$)  variation is more or less similar among all configurations. First, the velocity is constant in the inlet channel before the confluence region ($x/h<0$). There is  a  slight deceleration right after the confluence point at $x/h=0$ that is followed by a rapid acceleration before a high steady value at far downstream positions ($x/h\geq10$) is reached.
 However, a careful inspection of Fig.~\ref{fig:fig7_centrelinevelocity} and Fig.~\ref{fig:fig9_acceleration} unveils a considerable variation in the minimum and maximum centreline velocity, degree of deceleration and acceleration for different confluence angles $\beta$. These variations will now be studied in detail.

 Figure~\ref{fig:fig9_acceleration} shows the variation of minimum ($U_{c,min}$), maximum  ($U_{c,max}$) centreline velocity and strain rate $\dot{\varepsilon}$ ($\dv*{U}{x}$)
 along the centreline for all the  geometrical configurations. In the case of  the reference configuration ($\beta$ =~90$\degree$), close to the confluence point $x/h$~$\simeq$~0,  the deceleration is ascertained to be small as observed from Figs.~\ref{fig:fig7_centrelinevelocity}(a) (confluence region) and \ref{fig:fig9_acceleration}(c) (cyan color). As a consequence, the minimum centreline velocity $U_{c,min}$ is higher (meaning lower deceleration) than the other configurations as seen from Fig.~\ref{fig:fig9_acceleration}(a). Subsequent to deceleration, a substantial increase in the velocity caused by the sheath flow momentum normal to the core flow leads to acceleration of the core flow. As depicted in  Fig.~\ref{fig:fig9_acceleration}(c), $\dot{\varepsilon}$ (cyan color) continues to increase till the end of the confluence region ($x/h$~$\sim$~1), and  thereafter, $\dot{\varepsilon}$ decays to 0 at around $x/h$~$\approx$~5  since the velocity is almost constant as observed in Fig.~\ref{fig:fig7_centrelinevelocity}(a). However,  far downstream, as seen in Fig.~\ref{fig:fig9_acceleration}(b), $U_{c,max}$ is lowest as compared to all the other configurations. 
 
 For  the $\beta=[60$\degree$, 120$\degree$]$ pair, the velocity behaviour duplicate each other as observed from Figs.~\ref{fig:fig7_centrelinevelocity}(b) and \ref{fig:fig7_centrelinevelocity}(c) and more clearly from the strain rate $\dot{\varepsilon}$ plot in Fig.~\ref{fig:fig9_acceleration}(c). It is worth to recall that this similarity of the flow  mirrors the similarity of the thread shape evolution as seen earlier from Figs.~\ref{fig:fig4_Octthread}(b), \ref{fig:fig4_Octthread}(c) to Figs.~\ref{fig:fig7_Threadwidth_height}(b), \ref{fig:fig7_Threadwidth_height}(c).  In both cases, the effect of $\beta$ appends the streamwise flow through the contribution by sheath flow momemtum acting parallel to the core flow.  As a result, the core fluid velocity increases much faster than the reference configuration velocity. For both cases, $U_{c,min}$ and $U_{c,max}$ are symmetric  with respect to $\beta = 90\degree$  as seen in Figs.~\ref{fig:fig9_acceleration}(a) and ~\ref{fig:fig9_acceleration}(b). However, $U_{c,min}$ is lower for the $\beta = 90\degree$ case (implying higher deceleration) and $U_{c,max}$ is highest for  $[60$\degree$, 120$\degree$]$ cases when compared to all the other configruations. As observed from Fig.~\ref{fig:fig9_acceleration}(c), in both configurations, the variation in $\dot{\varepsilon}$ (golden and green color) shows a similar trend as the $\beta = 90\degree$ case (cyan color) but the magnitude is higher near the end of  the confluence regions ($x/h$~$\sim$~1.15) and from there onwards, the decay is much slower upto $x/h$~$\approx$~5.
 
 On the contrary,  Figs.~\ref{fig:fig7_centrelinevelocity}(d) and \ref{fig:fig7_centrelinevelocity}(e) display a quite different behaviour for  the $\beta=[30$\degree$, 150$\degree$]$ pair. Even though both the configurations have the same confluence region  0~$\leq$~$x/h$~$\leq$~2, there is a  considerable variation in the deceleration and acceleration regions as seen from Fig.~\ref{fig:fig9_acceleration}(c).  For  the $\beta=30\degree$ case, the deceleration ($x/h$~$\simeq$~0) is higher leading to the lowest $U_{c,min}$ among all the configurations and also lower $U_{c,max}$ far downstream as seen in Figs.~\ref{fig:fig9_acceleration}(a) and \ref{fig:fig9_acceleration}(b), whereas in the $\beta=150\degree$ case, deceleration is slightly lower as compared to the $\beta=30\degree$ case and $U_{c,max}$ is also higher. This results in an asymmetry of $U_{c,min}$ and $U_{c,max}$ with respect to the $\beta = 90\degree$ case.
 As seen from Fig.~\ref{fig:fig9_acceleration}(c) for $\beta=30\degree$, the strain rate $\dot{\varepsilon}$ (purple color) is more steep than for the $\beta=150\degree$ case leading to a lower peak value close to the end of confluence region ($x/h$~$\sim$~2). For $x/h > 2$ onwards, the decay of $\dot{\varepsilon}$ in both cases follows a similar trend. 

Thus, having affirmed the agreement between experimental measurements and numerical computations, we will now utilise purely the numerical computations for further analysis of the results in the sections that follow. In the next section, we look at the cross-sectional velocity distribution for all the geometrical configurations with varying $\beta$. \\

\begin{centering}
	\section{Cross-sectional velocity distribution} \label{cross-sectional velocities}
\end{centering}

\begin{figure} [tbp!]
	\centering
	\includegraphics[width=1\textwidth]{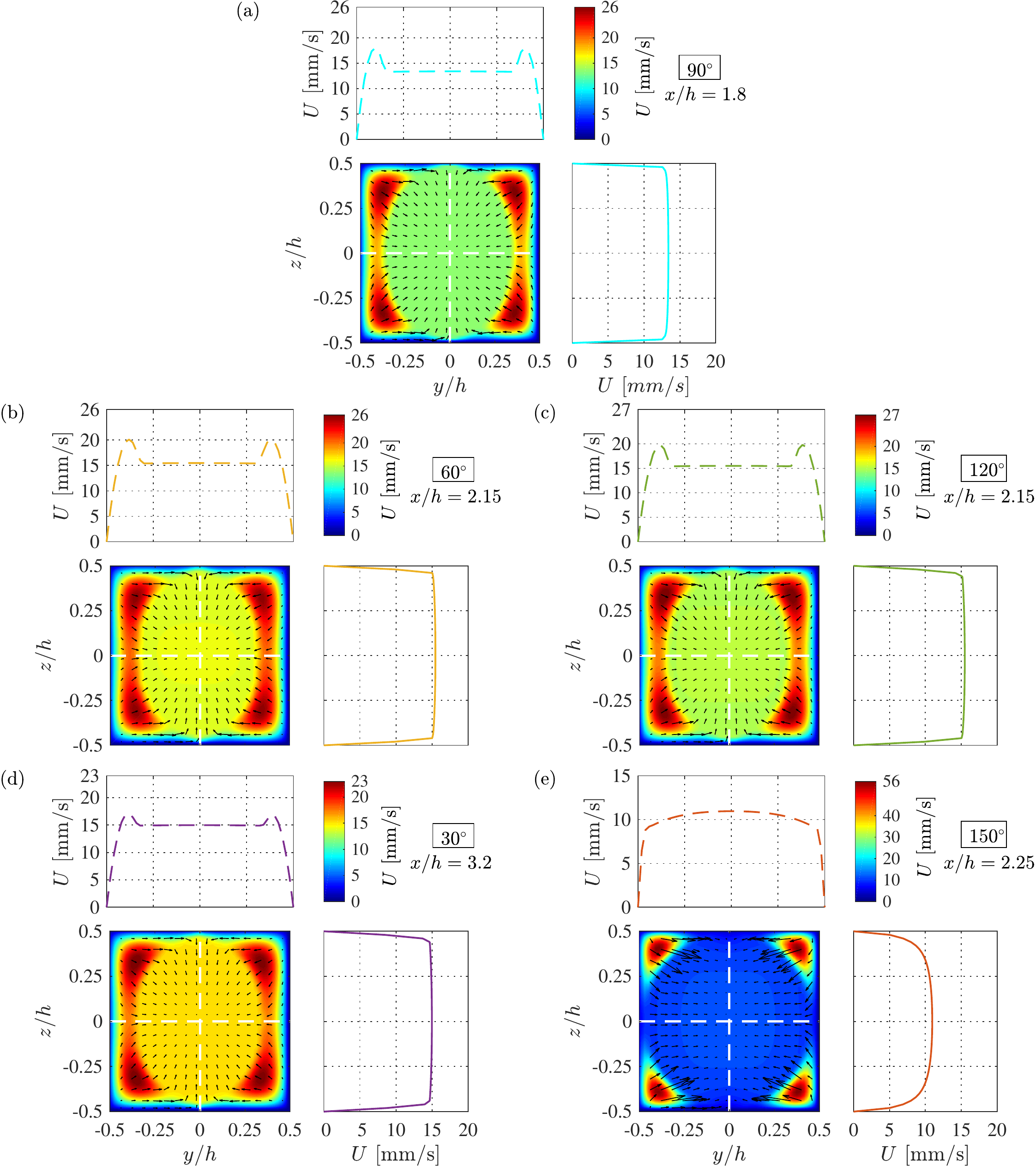} 
	\caption{(a) - (e) Cross-sectional velocity maps showing streamwise velocity  ($U$) superimposed with the in-plane velocity fields ($V$ and $W$, represented by black arrows) at different positions $x/h$ of flow-focusing configurations with varying  $\beta$. The positions $x/h$ correspond to the detachment locations of the core fluid  thread from the top wall namely the wetted lengths $L_w/h$ of respective configurations as per  Figs.~\ref{fig:fig5_Wetting_length}(a) - \ref{fig:fig5_Wetting_length}(e).  Velocity profiles denoted by solid and dashed lines represent the streamwise velocity component at $z/h, y/h$ = 0 (along the dashed white lines on the velocity maps). Note that the colorbar is different in each case (particularly  for  $\beta=150\degree$).}
	\label{fig10_velocitycontour}
\end{figure}

\begin{figure} [tbp!]
	\centering
	\includegraphics[width=1\textwidth]{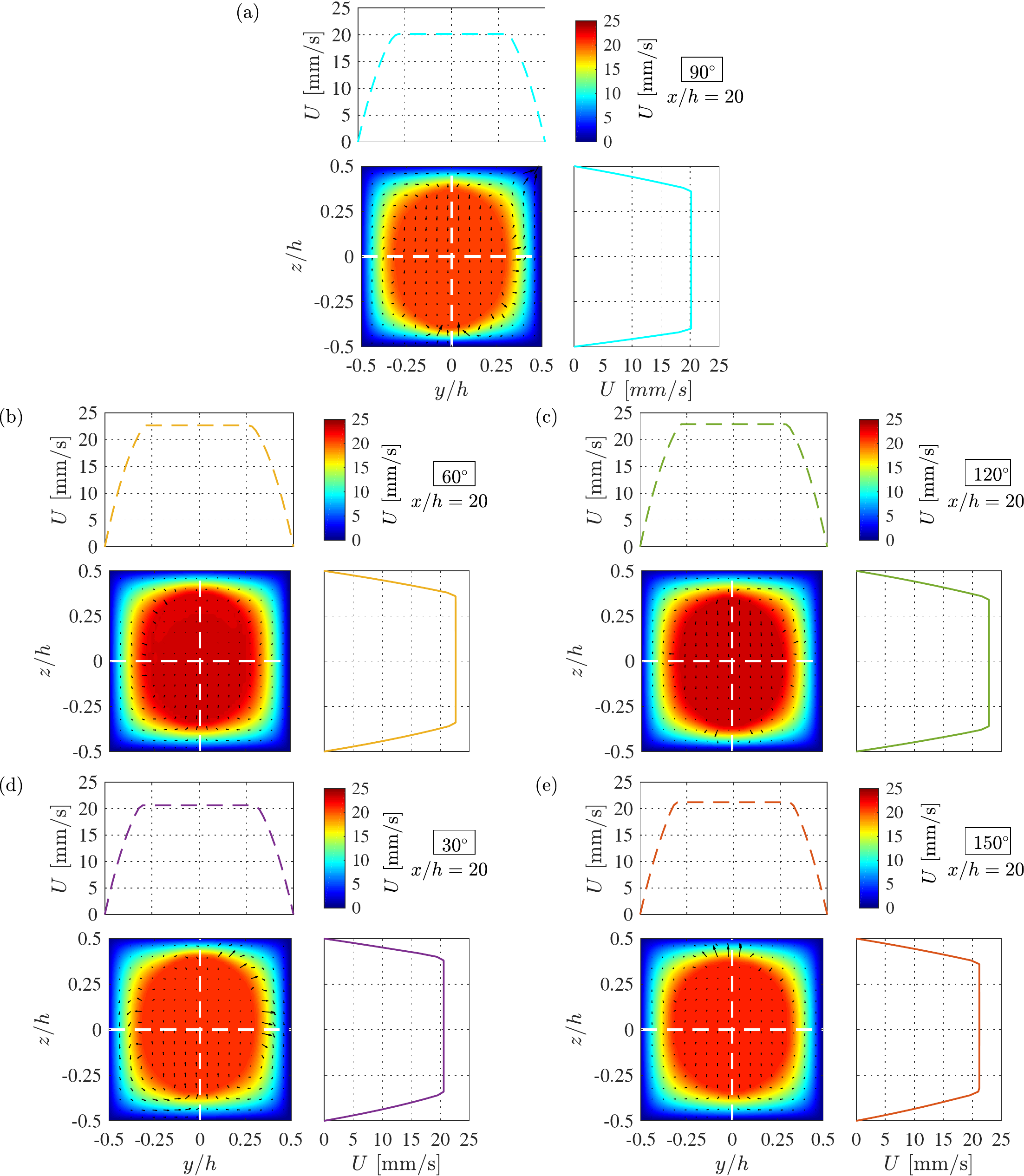} 
	\caption{(a) - (e) Cross-sectional velocity maps showing streamwise velocity ($U$) superimposed with in-plane velocity fields ($V$ and $W$,  represented by black arrows) at far downstream position $x/h$~=~20 for flow-focusing configurations with varying $\beta$. The in-plane velocities are very weak for all the configurations. Velocity  profiles of streamwise component are denoted by the solid and dashed lines at $z/h, y/h$ = 0 (demarcation shown by dashed white line on the velocity maps) show predominantly plug flow behaviour.}
	\label{fig11_velocitycontour}
\end{figure}

Figures~\ref{fig10_velocitycontour} and \ref{fig11_velocitycontour} show the numerical velocity maps superimposed with in-plane velocities ($V$ and $W$, 2-D vectors), together with the profiles of streamwise velocity component at $y/h$, $z/h$ = 0 in different cross-sectional $x/h$ planes for all the geometrical configurations. The cross-sectional $x/h$ planes in Fig.~\ref{fig10_velocitycontour} corresponds to the respective thread detachment position, i.e$.$ the wetted length $L_w/h$ of each configuration ($x/h\sim1.8$ for $\beta= 90\degree$; $x/h\sim 2.15$ for $\beta=[60$\degree$, 120$\degree$]$; $x/h\sim 3.2, 2.25$ for $\beta=[30$\degree$, 150$\degree$]$)  while in Fig.~\ref{fig11_velocitycontour},  $x/h$ planes  at  far downstream ($x/h=20$) are shown.

 As seen in Fig.~\ref{fig10_velocitycontour}, velocity maps depict the core fluid thread being enclosed by high-velocity sheath flows. Further, the 2-D vector plot of  in-plane velocities exhibit no evidence of appreciable recirculating/secondary flows at the corners in the cross-sectional  planes, albeit they are occupied by a region of  high-velocities. However, the magnitudes of these high-sheath velocities seem to differ significantly for different confluence angles $\beta$.  Accordingly, in the case of negative impingement configurations i.e. $\beta$ = 120$\degree$ and 150$\degree$, the velocity magnitudes are relatively higher and so is the length of 2-D vectors as seen in Figs.~\ref{fig10_velocitycontour}(c) and \ref{fig10_velocitycontour}(d).  They are maximum for the $\beta$ = 150$\degree$ case. The velocity profiles are nearly flat along the $z$-direction at $y/h=0$ in all configurations inside the core fluid thread.  Also, along the $y$-direction at $z/h=0$, a flat profile is observed  in the core with overshoots of  the sheath fluid velocity close to channel walls (with an exception for the $\beta$ = 150$\degree$ case).

Far downstream, as observed from Figs.~\ref{fig11_velocitycontour}(a) - \ref{fig11_velocitycontour}(e) for  $x/h=20$, the  velocity profiles within the colloidal thread stays uniform in both  the $y$ and $z$-directions signalling the plug flow behaviour.  Close examination of Figs.~\ref{fig11_velocitycontour}(b) and \ref{fig11_velocitycontour}(c)  display higher plug velocities for  the $\beta=[60$\degree$, 120$\degree$]$ pair amongst all the configurations, corroborating with earlier observations of maximum centreline velocity $U_{c,max}$ for the same cases in Fig.~\ref{fig:fig9_acceleration}(c). Furthermore, the 2-D vectors indicate the in-plane velocities for all the configurations to be very weak at this downstream position. Another  observation in all the configurations is that the velocity profiles near to the channel walls  are parabolic in both the $y$ and $z$-directions highlighting the sheath flows are wrapping the colloidal thread from all the four sides. \\

\begin{figure} [tbp!]
 	\centering
 	\includegraphics[width=1\textwidth]{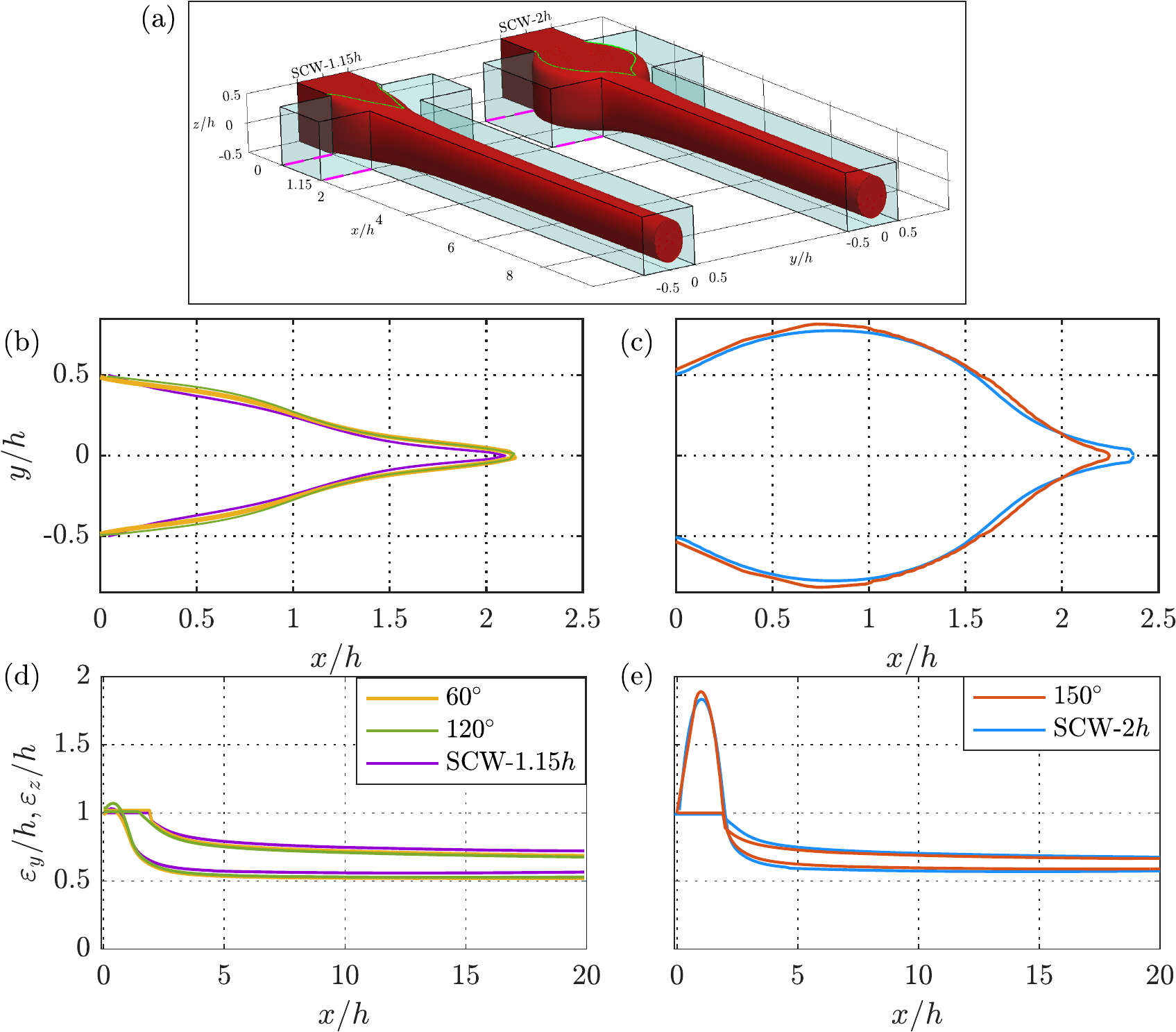} 
 	\caption{(a) 3-D view of  the thread shapes for flow-focusing configurations with varying sheath flow  channel widths (denoted by dashed magenta lines). Width of the side channel  in the `SCW-$1.5h$' configuration corresponds to the confluence region (0 $\leq$ $x/h$ $\leq$ 1.15) of  the $\beta=[60$\degree$, 120$\degree$]$ pair (see Figs.~\ref{fig:fig4_Octthread}(b), \ref{fig:fig4_Octthread}(c)).  Likewise, width of the side channel in `SCW-$2h$'  configuration corresponds to the confluence region (0~$\leq$~$x/h$~$\leq$~2) of $\beta=[30$\degree$, 150$\degree$]$ pair (see Figs.~\ref{fig:fig4_Octthread}(d), \ref{fig:fig4_Octthread}(e)). Panels (b) and (d)  shows the `SCW-$1.5h$' configuration wetted region, thread width ($\varepsilon_y/h$) and height ($\varepsilon_z/h$) plotted as a function of downstream positions $x/h$ overlapped with the wetted region and thread dimensions of  $\beta=[60$\degree$, 120$\degree$]$ pair. Similar plots of panels (b) and (d) in panels (c) and (e) for the case of  `SCW-$2h$' and $\beta= 150\degree$ configurations.}
 	\label{fig12_duplication}
 \end{figure}

\begin{centering}
	\section{Replication of confluence angle effects} \label{duplicate}
\end{centering}
In this section, a strategical approach was attempted numerically to comprehend the confluence angle effects. Two computations were performed utilising the reference  flow-focusing configuration ($\beta$ =~90$\degree$) with varying sheath flow inlet channel widths as illustrated in Fig.~\ref{fig12_duplication}(a). Accordingly, the side channel width (SCW) of  the `SCW-$1.15h$' configuration was set to $1.15h$ matching the confluence region (0 $\leq$ $x/h$ $\leq$ 1.15) of  the $\beta=[60$\degree$, 120$\degree$]$  configurations. Similarly,  the `SCW-$2h$' configuration width was fixed to $2h$  tallying with the confluence region  (0 $\leq$ $x/h$ $\leq$ 2) of  the $\beta=[30$\degree$, 150$\degree$]$ configurations. Other parameters such as flow rates of the core~($Q_{1}$) and sheath~($Q_{2})$ flows, rheologies of fluids are held fixed as described in the Secs.~\ref{geom-setup} and \ref{fluids}. The  interfacial tension $\gamma=\Gamma_{e}=0.615$~mN$~$m$^{-1}$, and the cross-sectional width~\textit{h} of  the central inlet and outlet channel were unchanged. This means that the sheath flow channel inlets are now  rectangular in these cases.

 Astonishingly, the flow in the SCW-$1.15h$ configuration precisely replicates the morphological features of  the wetted region and thread topologies of  the $\beta=[60$\degree$, 120$\degree$]$ configurations as observed from Figs.~\ref{fig12_duplication}(b)~and \ref{fig12_duplication}(d).  Once again this highlights the symmetry between these two cases. Whereas, the SCW-$2h$ configuration captures and replicates the features of only the $\beta$=150$\degree$  configuration as depicted in Figs.~\ref{fig12_duplication}(c) and~\ref{fig12_duplication}(e) (remember the asymmetry between the $\beta=[30$\degree$, 150$\degree$]$ configurations).  
 
 In the three cases, `SCW-$1.15h$'  and $\beta=[60$\degree$, 120$\degree$]$, the sheath flow momentum normal to the core flow is attenuated by 13.4\% in relative to the reference configuration ($\beta$ =~90$\degree$). For  the `SCW-$2h$' and $\beta=[30$\degree$, 150$\degree$]$ pair, the attenuation is around 50\%.  Therefore, inclining the sheath flow channels at various confluence angles $\beta$ or setting the width of sheath flow channels of reference configuration to confluence region lengths, would essentially mean inducing the same magnitude of sheath flow momentum normal to the core flow. Thus, from these demonstrations, it is aptly clear that the magnitude of  the sheath flow momentum at the confluence junction is the primary factor in controlling the morphological features of the wetted region and thread topologies except for  near co-flow ($\beta$~=~30$\degree$) case.  \\

\begin{centering}
\section{A comment on effective interfacial tension and thread detachment} \label{implication}
\end{centering}

The comprehensive systematic demonstrations utilising geometrically varying flow-focusing setups in Sec.~\ref{thread shapes} together with the alternative approach in previous Sec.~\ref{duplicate} underscores the significant role of $\Gamma_{e}$. As noticed, for a set flow rate and specified rheologies of fluids,
the same value of EIT ($\gamma=\Gamma_{e}=0.615$~mN$~$m$^{-1}$) deduced using the reference flow-focusing configuration ($\beta$~=~90$\degree$), very well reproduces the 3-D flow characteristics for the respective geometrical configurations in the numerics. All the features such as the thread evolutions, morphologies of the wetted regions, and velocity fields showed good agreement with the experimental findings in Secs.~\ref{thread shapes} and \ref{centreline velocities}. In addition, the replication of 3-D flow features of various confluence angle geometries captured via the modified geometries ($\beta$~=~90$\degree$), numerically with the same interfacial tension, further substantiate the robustness of our numerical procedure.

An important and interesting aspect from the above illustrations is on the detachment of thread from the top and bottom channel walls. We find that, the above value of interfacial tension $\gamma$ in the numerical computations not only captured the wetted region shapes and wetted lengths $L_{w}/h$ accurately, but also steered the thread detachment from the channel walls. Irrespective of the geometrical configurations, we also observed through a separate set of simulations that are not displayed here, that the wetted length $L_{w}/h \rightarrow \infty$  if  $\gamma  \rightarrow 0$, implying no detachment of thread from the top and bottom channel walls. Therefore, presence of ultra-low interfacial tension $\gamma$ ($\Gamma_{e}$) is a decisive factor in order for the thread detachment to occur. Nevertheless, given the presence of high viscosity contrast ($\chi \simeq 4500$) between the colloidal dispersion (core fluid) and its solvent (sheath fluid), \citet{cubaud2012} predict that the process of thread detachment in miscible systems having large viscosity contrast~\cite{cubaud2009,cubaud2014},
could occur naturally through the phenomena of self-lubrication~\cite{joseph1984} depending on the fluid injection geometries and injection flow rates. According to this principle, the low-viscosity fluid enwraps the high-viscosity fluid to minimize the viscous dissipation of energy, leading to self-lubricated thread structures. However, as evident from the above observations, in high Péclet number microfluidic flows involving miscible systems with high-viscosity contrast, thread detachment is found to be clearly controlled by $\Gamma_{e}$ as well. Indeed, in order to have a finer understanding of the effects of EIT versus the ones associated with self-lubrication on the thread detachment, further detailed investigation is needed. For instance, variation of EIT, viscosity contrast and flow rate ratios, and how these parameters affect the  wetted length $L_{w}/h$ and thereby the thread detachment is worthwhile, and such investigations are beyond the scope of this present work.\\

\begin{centering}
\section{Conclusions}\label{conclusions}
\end{centering}

The effective interfacial tension $\Gamma_e$ acting between the colloidal dispersion and its own solvent  has been estimated utilising the experimentally measured 3-D spatial evolution of thread shape in the flow-focusing channel and an effective capillary number dependent master curve obtained from a simple scaling model~\cite{gowda2019effective}. We have verified from the 3-D numerical computations, that the  $\Gamma_e$ which minimizes the error between computed and experimental thread heights is close to the estimated $\Gamma_e$ utilising the above method. The numerical computations were able to capture our experimental observations with good quantitative and qualitative agreement (both in terms of flow-topologies and flow-fields) for a range of confluence angles.  By mapping the numerical observations with the experimental measurements, we gained insights into the influence of various physical mechanisms driving the process of thread detachment from the top and bottom walls in the channels. From these meticulous analyses, we find that in the case of  high viscosity contrast miscible systems~\cite{cubaud2009,cubaud2014}, the thread detachment occuring in the physical experiments of microchannels may not be entirely based on the phenomena of self-lubrication~\cite{cubaud2012}  related with effect of viscous dissipation of energy,  but also the effect of  EIT induced by composition gradients plays  a major role. 

Moreover, these ultra-low interfacial tensions in non-equilibrium miscible fluids are extremely difficult to evaluate in  experimental measurements. The standard experimental techniques yielding reasonable results in the case of molecular miscible fluids are light scattering and spinning drop tensiometry~\cite{truzzolillo2017}. However, these two techniques are observed to have drawbacks, in particular at the transition region due to mixing of  the two fluids and difficulty to distinguish between the bulk fluids and interfacial region~\cite{truzzolillo2017}. The recently explored new measurement strategies like examining the Saffman–Taylor instability in a Hele-Shaw cell~\cite{truzzolillo2014,truzzolillo2016} with miscible complex fluids, and studying the dynamics of drop shape in spinning drop tensiometry~\cite{carbonaro2020} tested with miscible molecular systems are noteworthy developments. In addition to the above strategies, our method is also promising and could be a suitable choice.

 From a technical perspective, varying the effect of confluence angle $\beta$ assists in the utilisation of sheath flows  to  generate effective extensional flows. This, in turn, could facilitate to achieve optimal alignment of nanofibrils in colloidal dispersions~\cite{Brouzet_2018, Brouzet_2019}. A detailed investigation on the orientation and alignment of nanofibrils can be pursued, utilising the velocity gradients of various flow configurations, including  other relevant factors such as rotational diffusion, mobility, and  rigidity of fibrils. Concluding, a thorough understanding of effects like EIT  in miscible systems  opens up possibilities in tuning, and controlling the material properties more efficiently via appropriate selection of microfluidic geometrical configuration based on the practical interest and application. \\

\begin{acknowledgements} 
	Financial support by the Swedish Research Council for Environment,
Agricultural Sciences and Spatial Planning (FORMAS) and Swedish Research Council (VR) are gratefully acknowledged. The simulations were performed on computer resources provided by  the  Swedish National Infrastructure for Computing (SNIC) at National  Supercomputer  Centre, Linköping and at the Centre for High Performance Computing (PDC), KTH Royal Institute of Technology, Stockholm. Dr.~Korneliya Gordeyeva is acknowledged for experimental assistance. The authors thank Dr.~Christophe Brouzet  for  helpful discussions.
\end{acknowledgements}

\bibliography{Manuscript.bib}

\begin{thebibliography}{59}%
\makeatletter
\providecommand \@ifxundefined [1]{%
 \@ifx{#1\undefined}
}%
\providecommand \@ifnum [1]{%
 \ifnum #1\expandafter \@firstoftwo
 \else \expandafter \@secondoftwo
 \fi
}%
\providecommand \@ifx [1]{%
 \ifx #1\expandafter \@firstoftwo
 \else \expandafter \@secondoftwo
 \fi
}%
\providecommand \natexlab [1]{#1}%
\providecommand \enquote  [1]{``#1''}%
\providecommand \bibnamefont  [1]{#1}%
\providecommand \bibfnamefont [1]{#1}%
\providecommand \citenamefont [1]{#1}%
\providecommand \href@noop [0]{\@secondoftwo}%
\providecommand \href [0]{\begingroup \@sanitize@url \@href}%
\providecommand \@href[1]{\@@startlink{#1}\@@href}%
\providecommand \@@href[1]{\endgroup#1\@@endlink}%
\providecommand \@sanitize@url [0]{\catcode `\\12\catcode `\$12\catcode
  `\&12\catcode `\#12\catcode `\^12\catcode `\_12\catcode `\%12\relax}%
\providecommand \@@startlink[1]{}%
\providecommand \@@endlink[0]{}%
\providecommand \url  [0]{\begingroup\@sanitize@url \@url }%
\providecommand \@url [1]{\endgroup\@href {#1}{\urlprefix }}%
\providecommand \urlprefix  [0]{URL }%
\providecommand \Eprint [0]{\href }%
\providecommand \doibase [0]{https://doi.org/}%
\providecommand \selectlanguage [0]{\@gobble}%
\providecommand \bibinfo  [0]{\@secondoftwo}%
\providecommand \bibfield  [0]{\@secondoftwo}%
\providecommand \translation [1]{[#1]}%
\providecommand \BibitemOpen [0]{}%
\providecommand \bibitemStop [0]{}%
\providecommand \bibitemNoStop [0]{.\EOS\space}%
\providecommand \EOS [0]{\spacefactor3000\relax}%
\providecommand \BibitemShut  [1]{\csname bibitem#1\endcsname}%
\let\auto@bib@innerbib\@empty
\bibitem [{\citenamefont {Joseph}(1990)}]{joseph1990fluid}%
  \BibitemOpen
  \bibfield  {author} {\bibinfo {author} {\bibfnamefont {D.~D.}\ \bibnamefont
  {Joseph}},\ }\bibfield  {title} {\bibinfo {title} {Fluid dynamics of two
  miscible liquids with diffusion and gradient stresses},\ }\href@noop {}
  {\bibfield  {journal} {\bibinfo  {journal} {Eur. J. Mech. B Fluids}\ }\textbf
  {\bibinfo {volume} {9}},\ \bibinfo {pages} {565} (\bibinfo {year}
  {1990})}\BibitemShut {NoStop}%
\bibitem [{\citenamefont {Garik}\ \emph {et~al.}(1991)\citenamefont {Garik},
  \citenamefont {Hetrick}, \citenamefont {Orr}, \citenamefont {Barkey},\ and\
  \citenamefont {Ben-Jacob}}]{garik1991}%
  \BibitemOpen
  \bibfield  {author} {\bibinfo {author} {\bibfnamefont {P.}~\bibnamefont
  {Garik}}, \bibinfo {author} {\bibfnamefont {J.}~\bibnamefont {Hetrick}},
  \bibinfo {author} {\bibfnamefont {B.}~\bibnamefont {Orr}}, \bibinfo {author}
  {\bibfnamefont {D.}~\bibnamefont {Barkey}},\ and\ \bibinfo {author}
  {\bibfnamefont {E.}~\bibnamefont {Ben-Jacob}},\ }\bibfield  {title} {\bibinfo
  {title} {Interfacial cellular mixing and a conjecture on global deposit
  morphology},\ }\href@noop {} {\bibfield  {journal} {\bibinfo  {journal}
  {Phys. Rev. Lett.}\ }\textbf {\bibinfo {volume} {66}},\ \bibinfo {pages}
  {1606} (\bibinfo {year} {1991})}\BibitemShut {NoStop}%
\bibitem [{\citenamefont {Joseph}\ and\ \citenamefont
  {Renardy}(1993)}]{joseph1993fundamentals}%
  \BibitemOpen
  \bibfield  {author} {\bibinfo {author} {\bibfnamefont {D.~D.}\ \bibnamefont
  {Joseph}}\ and\ \bibinfo {author} {\bibfnamefont {Y.~Y.}\ \bibnamefont
  {Renardy}},\ }\href@noop {} {\emph {\bibinfo {title} {Fundamentals of
  Two-Fluid Dynamics. Part II: Lubricated Transport, Drops and Miscible
  Liquids}}}\ (\bibinfo  {publisher} {Springer, New York},\ \bibinfo {year}
  {1993})\BibitemShut {NoStop}%
\bibitem [{\citenamefont {Anderson}\ \emph {et~al.}(1998)\citenamefont
  {Anderson}, \citenamefont {McFadden},\ and\ \citenamefont
  {Wheeler}}]{anderson1998}%
  \BibitemOpen
  \bibfield  {author} {\bibinfo {author} {\bibfnamefont {D.~M.}\ \bibnamefont
  {Anderson}}, \bibinfo {author} {\bibfnamefont {G.~B.}\ \bibnamefont
  {McFadden}},\ and\ \bibinfo {author} {\bibfnamefont {A.~A.}\ \bibnamefont
  {Wheeler}},\ }\bibfield  {title} {\bibinfo {title} {Diffuse-interface methods
  in fluid mechanics},\ }\href@noop {} {\bibfield  {journal} {\bibinfo
  {journal} {Annual Review of Fluid Mechanics}\ }\textbf {\bibinfo {volume}
  {30}},\ \bibinfo {pages} {139} (\bibinfo {year} {1998})}\BibitemShut
  {NoStop}%
\bibitem [{\citenamefont {Atencia}\ and\ \citenamefont
  {Beebe}(2005)}]{atencia2004}%
  \BibitemOpen
  \bibfield  {author} {\bibinfo {author} {\bibfnamefont {J.}~\bibnamefont
  {Atencia}}\ and\ \bibinfo {author} {\bibfnamefont {D.~J.}\ \bibnamefont
  {Beebe}},\ }\bibfield  {title} {\bibinfo {title} {Controlled microfluidic
  interfaces},\ }\href@noop {} {\bibfield  {journal} {\bibinfo  {journal}
  {Nature}\ }\textbf {\bibinfo {volume} {437}},\ \bibinfo {pages} {648}
  (\bibinfo {year} {2005})}\BibitemShut {NoStop}%
\bibitem [{\citenamefont {Korteweg}(1901)}]{Korteweg1901}%
  \BibitemOpen
  \bibfield  {author} {\bibinfo {author} {\bibfnamefont {D.~J.}\ \bibnamefont
  {Korteweg}},\ }\bibfield  {title} {\bibinfo {title} {{Sur la forme que
  prennent les \'equations du mouvement des fluides si l'on tient compte des
  forces capillaires caus\'ees par des variations de densit\'e}},\ }\href@noop
  {} {\bibfield  {journal} {\bibinfo  {journal} {Arch. N\'eerlandaises Sci.
  Exactes Naturelles}\ }\textbf {\bibinfo {volume} {6}} (\bibinfo {year}
  {1901})}\BibitemShut {NoStop}%
\bibitem [{\citenamefont {Truzzolillo}\ \emph {et~al.}(2014)\citenamefont
  {Truzzolillo}, \citenamefont {Mora}, \citenamefont {Dupas},\ and\
  \citenamefont {Cipelletti}}]{truzzolillo2014}%
  \BibitemOpen
  \bibfield  {author} {\bibinfo {author} {\bibfnamefont {D.}~\bibnamefont
  {Truzzolillo}}, \bibinfo {author} {\bibfnamefont {S.}~\bibnamefont {Mora}},
  \bibinfo {author} {\bibfnamefont {C.}~\bibnamefont {Dupas}},\ and\ \bibinfo
  {author} {\bibfnamefont {L.}~\bibnamefont {Cipelletti}},\ }\bibfield  {title}
  {\bibinfo {title} {Off-equilibrium surface tension in colloidal
  suspensions},\ }\href@noop {} {\bibfield  {journal} {\bibinfo  {journal}
  {Phys. Rev. Lett.}\ }\textbf {\bibinfo {volume} {112}},\ \bibinfo {pages}
  {128303} (\bibinfo {year} {2014})}\BibitemShut {NoStop}%
\bibitem [{\citenamefont {Truzzolillo}\ \emph {et~al.}(2016)\citenamefont
  {Truzzolillo}, \citenamefont {Mora}, \citenamefont {Dupas},\ and\
  \citenamefont {Cipelletti}}]{truzzolillo2016}%
  \BibitemOpen
  \bibfield  {author} {\bibinfo {author} {\bibfnamefont {D.}~\bibnamefont
  {Truzzolillo}}, \bibinfo {author} {\bibfnamefont {S.}~\bibnamefont {Mora}},
  \bibinfo {author} {\bibfnamefont {C.}~\bibnamefont {Dupas}},\ and\ \bibinfo
  {author} {\bibfnamefont {L.}~\bibnamefont {Cipelletti}},\ }\bibfield  {title}
  {\bibinfo {title} {Nonequilibrium interfacial tension in simple and complex
  fluids},\ }\href@noop {} {\bibfield  {journal} {\bibinfo  {journal} {Phys.
  Rev. X}\ }\textbf {\bibinfo {volume} {6}},\ \bibinfo {pages} {041057}
  (\bibinfo {year} {2016})}\BibitemShut {NoStop}%
\bibitem [{\citenamefont {Bessonov}\ \emph {et~al.}(2005)\citenamefont
  {Bessonov}, \citenamefont {Volpert}, \citenamefont {Pojman},\ and\
  \citenamefont {Zoltowski}}]{bessonov2005}%
  \BibitemOpen
  \bibfield  {author} {\bibinfo {author} {\bibfnamefont {N.}~\bibnamefont
  {Bessonov}}, \bibinfo {author} {\bibfnamefont {V.}~\bibnamefont {Volpert}},
  \bibinfo {author} {\bibfnamefont {J.~A.}\ \bibnamefont {Pojman}},\ and\
  \bibinfo {author} {\bibfnamefont {B.}~\bibnamefont {Zoltowski}},\ }\bibfield
  {title} {\bibinfo {title} {Numerical simulations of convection induced by
  {Korteweg} stresses in miscible polymer-monomer systems},\ }\href@noop {}
  {\bibfield  {journal} {\bibinfo  {journal} {Microgravity-Science and
  Technology}\ }\textbf {\bibinfo {volume} {17}},\ \bibinfo {pages} {8}
  (\bibinfo {year} {2005})}\BibitemShut {NoStop}%
\bibitem [{\citenamefont {Chen}\ \emph {et~al.}(2001)\citenamefont {Chen},
  \citenamefont {Wang},\ and\ \citenamefont {Meiburg}}]{chen2001}%
  \BibitemOpen
  \bibfield  {author} {\bibinfo {author} {\bibfnamefont {C.}~\bibnamefont
  {Chen}}, \bibinfo {author} {\bibfnamefont {L.}~\bibnamefont {Wang}},\ and\
  \bibinfo {author} {\bibfnamefont {E.}~\bibnamefont {Meiburg}},\ }\bibfield
  {title} {\bibinfo {title} {Miscible droplets in a porous medium and the
  effects of {Korteweg} stresses},\ }\href@noop {} {\bibfield  {journal}
  {\bibinfo  {journal} {Phys. Fluids}\ }\textbf {\bibinfo {volume} {13}},\
  \bibinfo {pages} {2447} (\bibinfo {year} {2001})}\BibitemShut {NoStop}%
\bibitem [{\citenamefont {Chen}\ and\ \citenamefont
  {Meiburg}(2002)}]{chen2002}%
  \BibitemOpen
  \bibfield  {author} {\bibinfo {author} {\bibfnamefont {C.}~\bibnamefont
  {Chen}}\ and\ \bibinfo {author} {\bibfnamefont {E.}~\bibnamefont {Meiburg}},\
  }\bibfield  {title} {\bibinfo {title} {Miscible displacements in capillary
  tubes: Influence of {Korteweg} stresses and divergence effects},\ }\href@noop
  {} {\bibfield  {journal} {\bibinfo  {journal} {Phys. Fluids}\ }\textbf
  {\bibinfo {volume} {14}},\ \bibinfo {pages} {2052} (\bibinfo {year}
  {2002})}\BibitemShut {NoStop}%
\bibitem [{\citenamefont {Chen}\ \emph {et~al.}(2008)\citenamefont {Chen},
  \citenamefont {Huang}, \citenamefont {Gad{\^e}lha},\ and\ \citenamefont
  {Miranda}}]{chen2008}%
  \BibitemOpen
  \bibfield  {author} {\bibinfo {author} {\bibfnamefont {C.}~\bibnamefont
  {Chen}}, \bibinfo {author} {\bibfnamefont {C.}~\bibnamefont {Huang}},
  \bibinfo {author} {\bibfnamefont {H.}~\bibnamefont {Gad{\^e}lha}},\ and\
  \bibinfo {author} {\bibfnamefont {J.}~\bibnamefont {Miranda}},\ }\bibfield
  {title} {\bibinfo {title} {Radial viscous fingering in miscible {Hele}-{Shaw}
  flows: A numerical study},\ }\href@noop {} {\bibfield  {journal} {\bibinfo
  {journal} {Phys. Rev. E}\ }\textbf {\bibinfo {volume} {78}},\ \bibinfo
  {pages} {016306} (\bibinfo {year} {2008})}\BibitemShut {NoStop}%
\bibitem [{\citenamefont {Dias}\ and\ \citenamefont
  {Miranda}(2013)}]{dias2013}%
  \BibitemOpen
  \bibfield  {author} {\bibinfo {author} {\bibfnamefont {E.~O.}\ \bibnamefont
  {Dias}}\ and\ \bibinfo {author} {\bibfnamefont {J.~A.}\ \bibnamefont
  {Miranda}},\ }\bibfield  {title} {\bibinfo {title} {Wavelength selection in
  {Hele}-{Shaw} flows: A maximum-amplitude criterion},\ }\href@noop {}
  {\bibfield  {journal} {\bibinfo  {journal} {Phys. Rev. E}\ }\textbf {\bibinfo
  {volume} {88}},\ \bibinfo {pages} {013016} (\bibinfo {year}
  {2013})}\BibitemShut {NoStop}%
\bibitem [{\citenamefont {Mossige}\ \emph {et~al.}(2020)\citenamefont
  {Mossige}, \citenamefont {S.V.}, \citenamefont {Islamov}, \citenamefont
  {Wheeler},\ and\ \citenamefont {Fuller}}]{mossige2020}%
  \BibitemOpen
  \bibfield  {author} {\bibinfo {author} {\bibfnamefont {E.}~\bibnamefont
  {Mossige}}, \bibinfo {author} {\bibfnamefont {C.}~\bibnamefont {S.V.}},
  \bibinfo {author} {\bibfnamefont {M.}~\bibnamefont {Islamov}}, \bibinfo
  {author} {\bibfnamefont {S.}~\bibnamefont {Wheeler}},\ and\ \bibinfo {author}
  {\bibfnamefont {G.~G.}\ \bibnamefont {Fuller}},\ }\bibfield  {title}
  {\bibinfo {title} {Evaporation-induced {Rayleigh}--{Taylor} instabilities in
  polymer solutions},\ }\href@noop {} {\bibfield  {journal} {\bibinfo
  {journal} {Philosophical Transactions of the Royal Society A}\ }\textbf
  {\bibinfo {volume} {378}},\ \bibinfo {pages} {20190533} (\bibinfo {year}
  {2020})}\BibitemShut {NoStop}%
\bibitem [{\citenamefont {Pojmanm}\ \emph {et~al.}(2006)\citenamefont
  {Pojmanm}, \citenamefont {Whitmore}, \citenamefont {Turco}, \citenamefont
  {Maria}, \citenamefont {Lombardo}, \citenamefont {Marszalek}, \citenamefont
  {Parker},\ and\ \citenamefont {Zoltowski}}]{pojman2006}%
  \BibitemOpen
  \bibfield  {author} {\bibinfo {author} {\bibfnamefont {J.~A.}\ \bibnamefont
  {Pojmanm}}, \bibinfo {author} {\bibfnamefont {C.}~\bibnamefont {Whitmore}},
  \bibinfo {author} {\bibfnamefont {L.}~\bibnamefont {Turco}}, \bibinfo
  {author} {\bibfnamefont {L.}~\bibnamefont {Maria}}, \bibinfo {author}
  {\bibfnamefont {R.}~\bibnamefont {Lombardo}}, \bibinfo {author}
  {\bibfnamefont {J.}~\bibnamefont {Marszalek}}, \bibinfo {author}
  {\bibfnamefont {R.}~\bibnamefont {Parker}},\ and\ \bibinfo {author}
  {\bibfnamefont {B.}~\bibnamefont {Zoltowski}},\ }\bibfield  {title} {\bibinfo
  {title} {Evidence for the existence of an effective interfacial tension
  between miscible fluids: Isobutyric acid- water and 1-butanol- water in a
  spinning-drop tensiometer},\ }\href@noop {} {\bibfield  {journal} {\bibinfo
  {journal} {Langmuir}\ }\textbf {\bibinfo {volume} {22}},\ \bibinfo {pages}
  {2569} (\bibinfo {year} {2006})}\BibitemShut {NoStop}%
\bibitem [{\citenamefont {Carbonaro}\ \emph {et~al.}(2020)\citenamefont
  {Carbonaro}, \citenamefont {Cipelletti},\ and\ \citenamefont
  {Truzzolillo}}]{carbonaro2020}%
  \BibitemOpen
  \bibfield  {author} {\bibinfo {author} {\bibfnamefont {A.}~\bibnamefont
  {Carbonaro}}, \bibinfo {author} {\bibfnamefont {L.}~\bibnamefont
  {Cipelletti}},\ and\ \bibinfo {author} {\bibfnamefont {D.}~\bibnamefont
  {Truzzolillo}},\ }\bibfield  {title} {\bibinfo {title} {Ultralow effective
  interfacial tension between miscible molecular fluids},\ }\href@noop {}
  {\bibfield  {journal} {\bibinfo  {journal} {Phys. Rev. Fluids}\ }\textbf
  {\bibinfo {volume} {5}},\ \bibinfo {pages} {074001} (\bibinfo {year}
  {2020})}\BibitemShut {NoStop}%
\bibitem [{\citenamefont {Shevtsova}\ \emph {et~al.}(2016)\citenamefont
  {Shevtsova}, \citenamefont {Gaponenko}, \citenamefont {Yasnou}, \citenamefont
  {Mialdun},\ and\ \citenamefont {Nepomnyashchy}}]{shevtsova2016}%
  \BibitemOpen
  \bibfield  {author} {\bibinfo {author} {\bibfnamefont {V.}~\bibnamefont
  {Shevtsova}}, \bibinfo {author} {\bibfnamefont {Y.}~\bibnamefont
  {Gaponenko}}, \bibinfo {author} {\bibfnamefont {V.}~\bibnamefont {Yasnou}},
  \bibinfo {author} {\bibfnamefont {A.}~\bibnamefont {Mialdun}},\ and\ \bibinfo
  {author} {\bibfnamefont {A.}~\bibnamefont {Nepomnyashchy}},\ }\bibfield
  {title} {\bibinfo {title} {Two-scale wave patterns on a periodically excited
  miscible liquid--liquid interface},\ }\href@noop {} {\bibfield  {journal}
  {\bibinfo  {journal} {J. Fluid. Mech.}\ }\textbf {\bibinfo {volume} {795}},\
  \bibinfo {pages} {409} (\bibinfo {year} {2016})}\BibitemShut {NoStop}%
\bibitem [{\citenamefont {May}\ and\ \citenamefont {Maher}(1991)}]{may1991}%
  \BibitemOpen
  \bibfield  {author} {\bibinfo {author} {\bibfnamefont {S.}~\bibnamefont
  {May}}\ and\ \bibinfo {author} {\bibfnamefont {J.}~\bibnamefont {Maher}},\
  }\bibfield  {title} {\bibinfo {title} {Capillary-wave relaxation for a
  meniscus between miscible liquids},\ }\href@noop {} {\bibfield  {journal}
  {\bibinfo  {journal} {Phys. Rev. Lett}\ }\textbf {\bibinfo {volume} {67}},\
  \bibinfo {pages} {2013} (\bibinfo {year} {1991})}\BibitemShut {NoStop}%
\bibitem [{\citenamefont {Cicuta}\ \emph {et~al.}(2001)\citenamefont {Cicuta},
  \citenamefont {Vailati},\ and\ \citenamefont {Giglio}}]{cicuta2001}%
  \BibitemOpen
  \bibfield  {author} {\bibinfo {author} {\bibfnamefont {P.}~\bibnamefont
  {Cicuta}}, \bibinfo {author} {\bibfnamefont {A.}~\bibnamefont {Vailati}},\
  and\ \bibinfo {author} {\bibfnamefont {M.}~\bibnamefont {Giglio}},\
  }\bibfield  {title} {\bibinfo {title} {Capillary-to-bulk crossover of
  nonequilibrium fluctuations in the free diffusion of a near-critical binary
  liquid mixture},\ }\href@noop {} {\bibfield  {journal} {\bibinfo  {journal}
  {Applied Optics}\ }\textbf {\bibinfo {volume} {40}},\ \bibinfo {pages} {4140}
  (\bibinfo {year} {2001})}\BibitemShut {NoStop}%
\bibitem [{\citenamefont {Truzzolillo}\ and\ \citenamefont
  {Cipelletti}(2017)}]{truzzolillo2017}%
  \BibitemOpen
  \bibfield  {author} {\bibinfo {author} {\bibfnamefont {D.}~\bibnamefont
  {Truzzolillo}}\ and\ \bibinfo {author} {\bibfnamefont {L.}~\bibnamefont
  {Cipelletti}},\ }\bibfield  {title} {\bibinfo {title} {Off-equilibrium
  surface tension in miscible fluids},\ }\href@noop {} {\bibfield  {journal}
  {\bibinfo  {journal} {Soft Matter}\ }\textbf {\bibinfo {volume} {13}},\
  \bibinfo {pages} {13} (\bibinfo {year} {2017})}\BibitemShut {NoStop}%
\bibitem [{\citenamefont {Gowda}\ \emph {et~al.}(2019)\citenamefont {Gowda},
  \citenamefont {Brouzet}, \citenamefont {Lefranc}, \citenamefont
  {S{\"o}derberg},\ and\ \citenamefont {Lundell}}]{gowda2019effective}%
  \BibitemOpen
  \bibfield  {author} {\bibinfo {author} {\bibfnamefont {K.}~\bibnamefont
  {Gowda}}, \bibinfo {author} {\bibfnamefont {C.}~\bibnamefont {Brouzet}},
  \bibinfo {author} {\bibfnamefont {T.}~\bibnamefont {Lefranc}}, \bibinfo
  {author} {\bibfnamefont {L.}~\bibnamefont {S{\"o}derberg}},\ and\ \bibinfo
  {author} {\bibfnamefont {F.}~\bibnamefont {Lundell}},\ }\bibfield  {title}
  {\bibinfo {title} {Effective interfacial tension in flow-focusing of
  colloidal dispersions: {3-D} numerical simulations and experiments},\
  }\href@noop {} {\bibfield  {journal} {\bibinfo  {journal} {J. Fluid. Mech.}\
  }\textbf {\bibinfo {volume} {876}},\ \bibinfo {pages} {1052} (\bibinfo {year}
  {2019})}\BibitemShut {NoStop}%
\bibitem [{\citenamefont {Cubaud}\ and\ \citenamefont
  {Mason}(2006)}]{cubaud2006}%
  \BibitemOpen
  \bibfield  {author} {\bibinfo {author} {\bibfnamefont {T.}~\bibnamefont
  {Cubaud}}\ and\ \bibinfo {author} {\bibfnamefont {T.}~\bibnamefont {Mason}},\
  }\bibfield  {title} {\bibinfo {title} {Folding of viscous threads in
  diverging microchannels},\ }\href@noop {} {\bibfield  {journal} {\bibinfo
  {journal} {Phys. Rev. Lett.}\ }\textbf {\bibinfo {volume} {96}},\ \bibinfo
  {pages} {114501} (\bibinfo {year} {2006})}\BibitemShut {NoStop}%
\bibitem [{\citenamefont {Cubaud}\ and\ \citenamefont
  {Mason}(2012)}]{cubaud2012}%
  \BibitemOpen
  \bibfield  {author} {\bibinfo {author} {\bibfnamefont {T.}~\bibnamefont
  {Cubaud}}\ and\ \bibinfo {author} {\bibfnamefont {T.}~\bibnamefont {Mason}},\
  }\bibfield  {title} {\bibinfo {title} {Interacting viscous instabilities in
  microfluidic systems},\ }\href@noop {} {\bibfield  {journal} {\bibinfo
  {journal} {Soft Matter}\ }\textbf {\bibinfo {volume} {8}},\ \bibinfo {pages}
  {10573} (\bibinfo {year} {2012})}\BibitemShut {NoStop}%
\bibitem [{\citenamefont {Bonhomme}\ \emph {et~al.}(2012)\citenamefont
  {Bonhomme}, \citenamefont {Leng},\ and\ \citenamefont
  {Colin}}]{bonhomme2012}%
  \BibitemOpen
  \bibfield  {author} {\bibinfo {author} {\bibfnamefont {O.}~\bibnamefont
  {Bonhomme}}, \bibinfo {author} {\bibfnamefont {J.}~\bibnamefont {Leng}},\
  and\ \bibinfo {author} {\bibfnamefont {A.}~\bibnamefont {Colin}},\ }\bibfield
   {title} {\bibinfo {title} {Microfluidic wet-spinning of alginate
  microfibers: A theoretical analysis of fiber formation},\ }\href@noop {}
  {\bibfield  {journal} {\bibinfo  {journal} {Soft Matter}\ }\textbf {\bibinfo
  {volume} {8}},\ \bibinfo {pages} {10641} (\bibinfo {year}
  {2012})}\BibitemShut {NoStop}%
\bibitem [{\citenamefont {Cubaud}\ and\ \citenamefont
  {Notaro}(2014)}]{cubaud2014}%
  \BibitemOpen
  \bibfield  {author} {\bibinfo {author} {\bibfnamefont {T.}~\bibnamefont
  {Cubaud}}\ and\ \bibinfo {author} {\bibfnamefont {S.}~\bibnamefont
  {Notaro}},\ }\bibfield  {title} {\bibinfo {title} {Regimes of miscible fluid
  thread formation in microfluidic focusing sections},\ }\href@noop {}
  {\bibfield  {journal} {\bibinfo  {journal} {Phys. Fluids}\ }\textbf {\bibinfo
  {volume} {26}},\ \bibinfo {pages} {122005} (\bibinfo {year}
  {2014})}\BibitemShut {NoStop}%
\bibitem [{\citenamefont {Cubaud}\ and\ \citenamefont
  {Mason}(2009)}]{cubaud2009}%
  \BibitemOpen
  \bibfield  {author} {\bibinfo {author} {\bibfnamefont {T.}~\bibnamefont
  {Cubaud}}\ and\ \bibinfo {author} {\bibfnamefont {T.~G.}\ \bibnamefont
  {Mason}},\ }\bibfield  {title} {\bibinfo {title} {High-viscosity fluid
  threads in weakly diffusive microfluidic systems},\ }\href@noop {} {\bibfield
   {journal} {\bibinfo  {journal} {New J. Phys.}\ }\textbf {\bibinfo {volume}
  {11}},\ \bibinfo {pages} {075029} (\bibinfo {year} {2009})}\BibitemShut
  {NoStop}%
\bibitem [{\citenamefont {Cubaud}(2020)}]{cubaud2020}%
  \BibitemOpen
  \bibfield  {author} {\bibinfo {author} {\bibfnamefont {T.}~\bibnamefont
  {Cubaud}},\ }\bibfield  {title} {\bibinfo {title} {Swelling of diffusive
  fluid threads in microchannels},\ }\href@noop {} {\bibfield  {journal}
  {\bibinfo  {journal} {Phys. Rev. Lett.}\ }\textbf {\bibinfo {volume} {125}},\
  \bibinfo {pages} {174502} (\bibinfo {year} {2020})}\BibitemShut {NoStop}%
\bibitem [{\citenamefont {Anna}\ \emph {et~al.}(2003)\citenamefont {Anna},
  \citenamefont {Bontoux},\ and\ \citenamefont {Stone}}]{anna2003}%
  \BibitemOpen
  \bibfield  {author} {\bibinfo {author} {\bibfnamefont {S.~L.}\ \bibnamefont
  {Anna}}, \bibinfo {author} {\bibfnamefont {N.}~\bibnamefont {Bontoux}},\ and\
  \bibinfo {author} {\bibfnamefont {H.~A.}\ \bibnamefont {Stone}},\ }\bibfield
  {title} {\bibinfo {title} {Formation of dispersions using flow focusing in
  microchannels},\ }\href@noop {} {\bibfield  {journal} {\bibinfo  {journal}
  {App. Phys. Lett.}\ }\textbf {\bibinfo {volume} {82}},\ \bibinfo {pages}
  {364} (\bibinfo {year} {2003})}\BibitemShut {NoStop}%
\bibitem [{\citenamefont {Garstecki}\ \emph {et~al.}(2004)\citenamefont
  {Garstecki}, \citenamefont {Gitlin}, \citenamefont {DiLuzio}, \citenamefont
  {Whitesides}, \citenamefont {Kumacheva},\ and\ \citenamefont
  {Stone}}]{garstecki2004}%
  \BibitemOpen
  \bibfield  {author} {\bibinfo {author} {\bibfnamefont {P.}~\bibnamefont
  {Garstecki}}, \bibinfo {author} {\bibfnamefont {I.}~\bibnamefont {Gitlin}},
  \bibinfo {author} {\bibfnamefont {W.}~\bibnamefont {DiLuzio}}, \bibinfo
  {author} {\bibfnamefont {G.~M.}\ \bibnamefont {Whitesides}}, \bibinfo
  {author} {\bibfnamefont {E.}~\bibnamefont {Kumacheva}},\ and\ \bibinfo
  {author} {\bibfnamefont {H.~A.}\ \bibnamefont {Stone}},\ }\bibfield  {title}
  {\bibinfo {title} {Formation of monodisperse bubbles in a microfluidic
  flow-focusing device},\ }\href@noop {} {\bibfield  {journal} {\bibinfo
  {journal} {App. Phys. Lett.}\ }\textbf {\bibinfo {volume} {85}},\ \bibinfo
  {pages} {2649} (\bibinfo {year} {2004})}\BibitemShut {NoStop}%
\bibitem [{\citenamefont {Whitesides}(2006)}]{whitesides2006}%
  \BibitemOpen
  \bibfield  {author} {\bibinfo {author} {\bibfnamefont {G.~M.}\ \bibnamefont
  {Whitesides}},\ }\bibfield  {title} {\bibinfo {title} {The origins and the
  future of microfluidics},\ }\href@noop {} {\bibfield  {journal} {\bibinfo
  {journal} {Nature}\ }\textbf {\bibinfo {volume} {442}},\ \bibinfo {pages}
  {368} (\bibinfo {year} {2006})}\BibitemShut {NoStop}%
\bibitem [{\citenamefont {Hirt}\ and\ \citenamefont
  {Nichols}(1981)}]{hirt1981volume}%
  \BibitemOpen
  \bibfield  {author} {\bibinfo {author} {\bibfnamefont {C.~W.}\ \bibnamefont
  {Hirt}}\ and\ \bibinfo {author} {\bibfnamefont {B.~D.}\ \bibnamefont
  {Nichols}},\ }\bibfield  {title} {\bibinfo {title} {Volume of fluid ({VoF})
  method for the dynamics of free boundaries},\ }\href@noop {} {\bibfield
  {journal} {\bibinfo  {journal} {J. Comput. Phys.}\ }\textbf {\bibinfo
  {volume} {39}},\ \bibinfo {pages} {201} (\bibinfo {year} {1981})}\BibitemShut
  {NoStop}%
\bibitem [{\citenamefont {OpenCFD}(2018)}]{OpenCFD}%
  \BibitemOpen
  \bibfield  {author} {\bibinfo {author} {\bibfnamefont {E.}~\bibnamefont
  {OpenCFD}},\ }\href@noop {} {\emph {\bibinfo {title} {OpenFOAM v1806
  Userguide}}}\ (\bibinfo  {publisher} {ESI-OpenCFD},\ \bibinfo {year}
  {2018})\BibitemShut {NoStop}%
\bibitem [{\citenamefont {Joseph}\ \emph {et~al.}(1984)\citenamefont {Joseph},
  \citenamefont {Nguyen},\ and\ \citenamefont {Beavers}}]{joseph1984}%
  \BibitemOpen
  \bibfield  {author} {\bibinfo {author} {\bibfnamefont {D.}~\bibnamefont
  {Joseph}}, \bibinfo {author} {\bibfnamefont {K.}~\bibnamefont {Nguyen}},\
  and\ \bibinfo {author} {\bibfnamefont {G.}~\bibnamefont {Beavers}},\
  }\bibfield  {title} {\bibinfo {title} {Non-uniqueness and stability of the
  configuration of flow of immiscible fluids with different viscosities},\
  }\href@noop {} {\bibfield  {journal} {\bibinfo  {journal} {J. Fluid. Mech.}\
  }\textbf {\bibinfo {volume} {141}},\ \bibinfo {pages} {319} (\bibinfo {year}
  {1984})}\BibitemShut {NoStop}%
\bibitem [{\citenamefont {H{\aa}kansson}\ \emph {et~al.}(2014)\citenamefont
  {H{\aa}kansson}, \citenamefont {Fall}, \citenamefont {Lundell}, \citenamefont
  {Yu}, \citenamefont {Krywka}, \citenamefont {Roth}, \citenamefont {Santoro},
  \citenamefont {Kvick}, \citenamefont {Prahl-Wittberg}, \citenamefont
  {W{\aa}gberg},\ and\ \citenamefont
  {S{\"o}derberg}}]{haakansson2014hydrodynamic}%
  \BibitemOpen
  \bibfield  {author} {\bibinfo {author} {\bibfnamefont {K.~M.~O.}\
  \bibnamefont {H{\aa}kansson}}, \bibinfo {author} {\bibfnamefont {A.~B.}\
  \bibnamefont {Fall}}, \bibinfo {author} {\bibfnamefont {F.}~\bibnamefont
  {Lundell}}, \bibinfo {author} {\bibfnamefont {S.}~\bibnamefont {Yu}},
  \bibinfo {author} {\bibfnamefont {C.}~\bibnamefont {Krywka}}, \bibinfo
  {author} {\bibfnamefont {S.~V.}\ \bibnamefont {Roth}}, \bibinfo {author}
  {\bibfnamefont {G.}~\bibnamefont {Santoro}}, \bibinfo {author} {\bibfnamefont
  {M.}~\bibnamefont {Kvick}}, \bibinfo {author} {\bibfnamefont
  {L.}~\bibnamefont {Prahl-Wittberg}}, \bibinfo {author} {\bibfnamefont
  {L.}~\bibnamefont {W{\aa}gberg}},\ and\ \bibinfo {author} {\bibfnamefont
  {L.~D.}\ \bibnamefont {S{\"o}derberg}},\ }\bibfield  {title} {\bibinfo
  {title} {Hydrodynamic alignment and assembly of nanofibrils resulting in
  strong cellulose filaments},\ }\href@noop {} {\bibfield  {journal} {\bibinfo
  {journal} {Nat. Commun.}\ }\textbf {\bibinfo {volume} {5}},\ \bibinfo {pages}
  {4018} (\bibinfo {year} {2014})}\BibitemShut {NoStop}%
\bibitem [{\citenamefont {Mittal}\ \emph {et~al.}(2018)\citenamefont {Mittal},
  \citenamefont {Ansari}, \citenamefont {K.}, \citenamefont {Brouzet},
  \citenamefont {Chen}, \citenamefont {Larsson}, \citenamefont {Roth},
  \citenamefont {Lundell}, \citenamefont {Wagberg}, \citenamefont {Kotov},\
  and\ \citenamefont {S{\"o}derberg}}]{mittal2018}%
  \BibitemOpen
  \bibfield  {author} {\bibinfo {author} {\bibfnamefont {N.}~\bibnamefont
  {Mittal}}, \bibinfo {author} {\bibfnamefont {F.}~\bibnamefont {Ansari}},
  \bibinfo {author} {\bibfnamefont {G.~V.}\ \bibnamefont {K.}}, \bibinfo
  {author} {\bibfnamefont {C.}~\bibnamefont {Brouzet}}, \bibinfo {author}
  {\bibfnamefont {P.}~\bibnamefont {Chen}}, \bibinfo {author} {\bibfnamefont
  {P.~T.}\ \bibnamefont {Larsson}}, \bibinfo {author} {\bibfnamefont {S.~V.}\
  \bibnamefont {Roth}}, \bibinfo {author} {\bibfnamefont {F.}~\bibnamefont
  {Lundell}}, \bibinfo {author} {\bibfnamefont {L.}~\bibnamefont {Wagberg}},
  \bibinfo {author} {\bibfnamefont {N.~A.}\ \bibnamefont {Kotov}},\ and\
  \bibinfo {author} {\bibfnamefont {L.~D.}\ \bibnamefont {S{\"o}derberg}},\
  }\bibfield  {title} {\bibinfo {title} {Multiscale control of nanocellulose
  assembly: Transferring remarkable nanoscale fibril mechanics to macroscale
  fibers},\ }\href@noop {} {\bibfield  {journal} {\bibinfo  {journal} {ACS
  Nano}\ }\textbf {\bibinfo {volume} {12}},\ \bibinfo {pages} {6378} (\bibinfo
  {year} {2018})}\BibitemShut {NoStop}%
\bibitem [{\citenamefont {Brouzet}\ \emph {et~al.}(2018)\citenamefont
  {Brouzet}, \citenamefont {Mittal}, \citenamefont {S{\"o}derberg},\ and\
  \citenamefont {Lundell}}]{Brouzet_2018}%
  \BibitemOpen
  \bibfield  {author} {\bibinfo {author} {\bibfnamefont {C.}~\bibnamefont
  {Brouzet}}, \bibinfo {author} {\bibfnamefont {N.}~\bibnamefont {Mittal}},
  \bibinfo {author} {\bibfnamefont {L.~D.}\ \bibnamefont {S{\"o}derberg}},\
  and\ \bibinfo {author} {\bibfnamefont {F.}~\bibnamefont {Lundell}},\
  }\bibfield  {title} {\bibinfo {title} {Size-dependent orientational dynamics
  of {Brownian} nanorods},\ }\href@noop {} {\bibfield  {journal} {\bibinfo
  {journal} {ACS Macro Letters}\ }\textbf {\bibinfo {volume} {7}},\ \bibinfo
  {pages} {1022} (\bibinfo {year} {2018})}\BibitemShut {NoStop}%
\bibitem [{\citenamefont {Brouzet}\ \emph {et~al.}(2019)\citenamefont
  {Brouzet}, \citenamefont {Mittal}, , \citenamefont {Lundell},\ and\
  \citenamefont {S{\"o}derberg}}]{Brouzet_2019}%
  \BibitemOpen
  \bibfield  {author} {\bibinfo {author} {\bibfnamefont {C.}~\bibnamefont
  {Brouzet}}, \bibinfo {author} {\bibfnamefont {N.}~\bibnamefont {Mittal}}, ,
  \bibinfo {author} {\bibfnamefont {F.}~\bibnamefont {Lundell}},\ and\ \bibinfo
  {author} {\bibfnamefont {L.~D.}\ \bibnamefont {S{\"o}derberg}},\ }\bibfield
  {title} {\bibinfo {title} {Characterizing the orientational and network
  dynamics of polydisperse nanofibers on the nanoscale},\ }\href@noop {}
  {\bibfield  {journal} {\bibinfo  {journal} {Macromolecules}\ }\textbf
  {\bibinfo {volume} {52}},\ \bibinfo {pages} {2286} (\bibinfo {year}
  {2019})}\BibitemShut {NoStop}%
\bibitem [{\citenamefont {Truzzolillo}\ \emph {et~al.}(2015)\citenamefont
  {Truzzolillo}, \citenamefont {Roger}, \citenamefont {Dupas}, \citenamefont
  {Mora},\ and\ \citenamefont {Cipelletti}}]{truzzolillo2015}%
  \BibitemOpen
  \bibfield  {author} {\bibinfo {author} {\bibfnamefont {D.}~\bibnamefont
  {Truzzolillo}}, \bibinfo {author} {\bibfnamefont {V.}~\bibnamefont {Roger}},
  \bibinfo {author} {\bibfnamefont {C.}~\bibnamefont {Dupas}}, \bibinfo
  {author} {\bibfnamefont {S.}~\bibnamefont {Mora}},\ and\ \bibinfo {author}
  {\bibfnamefont {L.}~\bibnamefont {Cipelletti}},\ }\bibfield  {title}
  {\bibinfo {title} {Bulk and interfacial stresses in suspensions of soft and
  hard colloids},\ }\href@noop {} {\bibfield  {journal} {\bibinfo  {journal}
  {Journal of Physics: Condensed Matter}\ }\textbf {\bibinfo {volume} {27}},\
  \bibinfo {pages} {194103} (\bibinfo {year} {2015})}\BibitemShut {NoStop}%
\bibitem [{\citenamefont {Saito}\ \emph {et~al.}(2007)\citenamefont {Saito},
  \citenamefont {Kimura}, \citenamefont {Nishiyama},\ and\ \citenamefont
  {Isogai}}]{saito2007}%
  \BibitemOpen
  \bibfield  {author} {\bibinfo {author} {\bibfnamefont {T.}~\bibnamefont
  {Saito}}, \bibinfo {author} {\bibfnamefont {S.}~\bibnamefont {Kimura}},
  \bibinfo {author} {\bibfnamefont {Y.}~\bibnamefont {Nishiyama}},\ and\
  \bibinfo {author} {\bibfnamefont {A.}~\bibnamefont {Isogai}},\ }\bibfield
  {title} {\bibinfo {title} {Cellulose nanofibers prepared by {Tempo}-mediated
  oxidation of native cellulose},\ }\href@noop {} {\bibfield  {journal}
  {\bibinfo  {journal} {Biomacromolecules}\ }\textbf {\bibinfo {volume} {8}},\
  \bibinfo {pages} {2485} (\bibinfo {year} {2007})}\BibitemShut {NoStop}%
\bibitem [{\citenamefont {Isogai}\ \emph {et~al.}(2011)\citenamefont {Isogai},
  \citenamefont {Saito},\ and\ \citenamefont {Fukuzumi}}]{isogai2011tempo}%
  \BibitemOpen
  \bibfield  {author} {\bibinfo {author} {\bibfnamefont {A.}~\bibnamefont
  {Isogai}}, \bibinfo {author} {\bibfnamefont {T.}~\bibnamefont {Saito}},\ and\
  \bibinfo {author} {\bibfnamefont {H.}~\bibnamefont {Fukuzumi}},\ }\bibfield
  {title} {\bibinfo {title} {Tempo-oxidized cellulose nanofibers},\ }\href@noop
  {} {\bibfield  {journal} {\bibinfo  {journal} {Nanoscale}\ }\textbf {\bibinfo
  {volume} {3}},\ \bibinfo {pages} {71} (\bibinfo {year} {2011})}\BibitemShut
  {NoStop}%
\bibitem [{\citenamefont {Marto{\"\i}a}\ \emph {et~al.}(2016)\citenamefont
  {Marto{\"\i}a}, \citenamefont {Dumont}, \citenamefont {Org{\'e}as},
  \citenamefont {Belgacem},\ and\ \citenamefont {Putaux}}]{martoia2016}%
  \BibitemOpen
  \bibfield  {author} {\bibinfo {author} {\bibfnamefont {F.}~\bibnamefont
  {Marto{\"\i}a}}, \bibinfo {author} {\bibfnamefont {P.~J.~J.}\ \bibnamefont
  {Dumont}}, \bibinfo {author} {\bibfnamefont {L.}~\bibnamefont {Org{\'e}as}},
  \bibinfo {author} {\bibfnamefont {M.~N.}\ \bibnamefont {Belgacem}},\ and\
  \bibinfo {author} {\bibfnamefont {J.~L.}\ \bibnamefont {Putaux}},\ }\bibfield
   {title} {\bibinfo {title} {Micro-mechanics of electrostatically stabilized
  suspensions of cellulose nanofibrils under steady state shear flow},\
  }\href@noop {} {\bibfield  {journal} {\bibinfo  {journal} {Soft matter}\
  }\textbf {\bibinfo {volume} {12}},\ \bibinfo {pages} {1721} (\bibinfo {year}
  {2016})}\BibitemShut {NoStop}%
\bibitem [{\citenamefont {Nechyporchuk}\ \emph {et~al.}(2016)\citenamefont
  {Nechyporchuk}, \citenamefont {Belgacem},\ and\ \citenamefont
  {Pignon}}]{Nechyporchuketal2016}%
  \BibitemOpen
  \bibfield  {author} {\bibinfo {author} {\bibfnamefont {O.}~\bibnamefont
  {Nechyporchuk}}, \bibinfo {author} {\bibfnamefont {M.~N.}\ \bibnamefont
  {Belgacem}},\ and\ \bibinfo {author} {\bibfnamefont {F.}~\bibnamefont
  {Pignon}},\ }\bibfield  {title} {\bibinfo {title} {Current progress in
  rheology of cellulose nanofibril suspensions},\ }\href@noop {} {\bibfield
  {journal} {\bibinfo  {journal} {Biomacromolecules}\ }\textbf {\bibinfo
  {volume} {17}},\ \bibinfo {pages} {2311} (\bibinfo {year}
  {2016})}\BibitemShut {NoStop}%
\bibitem [{\citenamefont {Geng}\ \emph {et~al.}(2018)\citenamefont {Geng},
  \citenamefont {Mittal}, \citenamefont {Zhan}, \citenamefont {Ansari},
  \citenamefont {Sharma}, \citenamefont {Peng}, \citenamefont {Hsiao},\ and\
  \citenamefont {S{\"o}derberg}}]{geng2018}%
  \BibitemOpen
  \bibfield  {author} {\bibinfo {author} {\bibfnamefont {L.}~\bibnamefont
  {Geng}}, \bibinfo {author} {\bibfnamefont {N.}~\bibnamefont {Mittal}},
  \bibinfo {author} {\bibfnamefont {C.}~\bibnamefont {Zhan}}, \bibinfo {author}
  {\bibfnamefont {F.}~\bibnamefont {Ansari}}, \bibinfo {author} {\bibfnamefont
  {P.~R.}\ \bibnamefont {Sharma}}, \bibinfo {author} {\bibfnamefont
  {X.}~\bibnamefont {Peng}}, \bibinfo {author} {\bibfnamefont {B.~S.}\
  \bibnamefont {Hsiao}},\ and\ \bibinfo {author} {\bibfnamefont
  {L.}~\bibnamefont {S{\"o}derberg}},\ }\bibfield  {title} {\bibinfo {title}
  {Understanding the mechanistic behavior of highly charged cellulose
  nanofibers in aqueous systems},\ }\href@noop {} {\bibfield  {journal}
  {\bibinfo  {journal} {Macromolecules}\ }\textbf {\bibinfo {volume} {51}},\
  \bibinfo {pages} {1498} (\bibinfo {year} {2018})}\BibitemShut {NoStop}%
\bibitem [{\citenamefont {Switzer}\ and\ \citenamefont
  {Klingenberg}(2003)}]{switzer2003}%
  \BibitemOpen
  \bibfield  {author} {\bibinfo {author} {\bibfnamefont {L.~H.}\ \bibnamefont
  {Switzer}}\ and\ \bibinfo {author} {\bibfnamefont {D.~J.}\ \bibnamefont
  {Klingenberg}},\ }\bibfield  {title} {\bibinfo {title} {Simulations of fiber
  floc dispersion in linear flow fields},\ }\href@noop {} {\bibfield  {journal}
  {\bibinfo  {journal} {Nordic Pulp \& Paper Research Journal}\ }\textbf
  {\bibinfo {volume} {18}},\ \bibinfo {pages} {141} (\bibinfo {year}
  {2003})}\BibitemShut {NoStop}%
\bibitem [{\citenamefont {Doi}\ and\ \citenamefont {Edwards}(1988)}]{doi1988}%
  \BibitemOpen
  \bibfield  {author} {\bibinfo {author} {\bibfnamefont {M.}~\bibnamefont
  {Doi}}\ and\ \bibinfo {author} {\bibfnamefont {S.~F.}\ \bibnamefont
  {Edwards}},\ }\href@noop {} {\emph {\bibinfo {title} {The theory of polymer
  dynamics}}},\ Vol.~\bibinfo {volume} {73}\ (\bibinfo  {publisher} {Oxford
  University Press},\ \bibinfo {year} {1988})\BibitemShut {NoStop}%
\bibitem [{\citenamefont {Marrucci}\ and\ \citenamefont
  {Grizzuti}(1983)}]{marrucci1983}%
  \BibitemOpen
  \bibfield  {author} {\bibinfo {author} {\bibfnamefont {G.}~\bibnamefont
  {Marrucci}}\ and\ \bibinfo {author} {\bibfnamefont {N.}~\bibnamefont
  {Grizzuti}},\ }\bibfield  {title} {\bibinfo {title} {The effect of
  polydispersity on rotational diffusivity and shear viscosity of rodlike
  polymers in concentrated solutions},\ }\href@noop {} {\bibfield  {journal}
  {\bibinfo  {journal} {Journal of Polymer Science: Polymer Letters Edition}\
  }\textbf {\bibinfo {volume} {21}},\ \bibinfo {pages} {83} (\bibinfo {year}
  {1983})}\BibitemShut {NoStop}%
\bibitem [{\citenamefont {Marrucci}\ and\ \citenamefont
  {Grizzuti}(1984)}]{marrucci1984}%
  \BibitemOpen
  \bibfield  {author} {\bibinfo {author} {\bibfnamefont {G.}~\bibnamefont
  {Marrucci}}\ and\ \bibinfo {author} {\bibfnamefont {N.}~\bibnamefont
  {Grizzuti}},\ }\bibfield  {title} {\bibinfo {title} {Predicted effect of
  polydispersity on rodlike polymer behaviour in concentrated solutions},\
  }\href@noop {} {\bibfield  {journal} {\bibinfo  {journal} {Journal of
  Non-Newtonian Fluid Mechanics}\ }\textbf {\bibinfo {volume} {14}},\ \bibinfo
  {pages} {103} (\bibinfo {year} {1984})}\BibitemShut {NoStop}%
\bibitem [{\citenamefont {Chow}\ \emph {et~al.}(1985)\citenamefont {Chow},
  \citenamefont {Fuller}, \citenamefont {Wallace},\ and\ \citenamefont
  {Madri}}]{chow1985}%
  \BibitemOpen
  \bibfield  {author} {\bibinfo {author} {\bibfnamefont {A.~W.}\ \bibnamefont
  {Chow}}, \bibinfo {author} {\bibfnamefont {G.~G.}\ \bibnamefont {Fuller}},
  \bibinfo {author} {\bibfnamefont {D.~G.}\ \bibnamefont {Wallace}},\ and\
  \bibinfo {author} {\bibfnamefont {J.~A.}\ \bibnamefont {Madri}},\ }\bibfield
  {title} {\bibinfo {title} {Rheo-optical response of rodlike, shortened
  collagen protein to transient shear flow},\ }\href@noop {} {\bibfield
  {journal} {\bibinfo  {journal} {Macromolecules}\ }\textbf {\bibinfo {volume}
  {18}},\ \bibinfo {pages} {805} (\bibinfo {year} {1985})}\BibitemShut
  {NoStop}%
\bibitem [{\citenamefont {Pecora}(1985)}]{pecora1985}%
  \BibitemOpen
  \bibfield  {author} {\bibinfo {author} {\bibfnamefont {R.}~\bibnamefont
  {Pecora}},\ }\bibfield  {title} {\bibinfo {title} {Dynamics of rodlike
  macromolecules in semidilute solutions},\ }\href@noop {} {\bibfield
  {journal} {\bibinfo  {journal} {Journal of Polymer Science: Polymer
  Symposia}\ }\textbf {\bibinfo {volume} {73}},\ \bibinfo {pages} {83}
  (\bibinfo {year} {1985})}\BibitemShut {NoStop}%
\bibitem [{\citenamefont {Rogers}\ \emph {et~al.}(2005)\citenamefont {Rogers},
  \citenamefont {Venema}, \citenamefont {Sagis}, \citenamefont {van~der
  Linden},\ and\ \citenamefont {Donald}}]{rogers2005}%
  \BibitemOpen
  \bibfield  {author} {\bibinfo {author} {\bibfnamefont {S.~S.}\ \bibnamefont
  {Rogers}}, \bibinfo {author} {\bibfnamefont {P.}~\bibnamefont {Venema}},
  \bibinfo {author} {\bibfnamefont {L.~M.~C.}\ \bibnamefont {Sagis}}, \bibinfo
  {author} {\bibfnamefont {E.}~\bibnamefont {van~der Linden}},\ and\ \bibinfo
  {author} {\bibfnamefont {A.~M.}\ \bibnamefont {Donald}},\ }\bibfield  {title}
  {\bibinfo {title} {Measuring the length distribution of a fibril system: a
  flow birefringence technique applied to amyloid fibrils},\ }\href@noop {}
  {\bibfield  {journal} {\bibinfo  {journal} {Macromolecules}\ }\textbf
  {\bibinfo {volume} {38}},\ \bibinfo {pages} {2948} (\bibinfo {year}
  {2005})}\BibitemShut {NoStop}%
\bibitem [{\citenamefont {Ros{\'e}n}\ \emph {et~al.}(2020)\citenamefont
  {Ros{\'e}n}, \citenamefont {Hsiao},\ and\ \citenamefont
  {S{\"o}derberg}}]{rosen2020}%
  \BibitemOpen
  \bibfield  {author} {\bibinfo {author} {\bibfnamefont {T.}~\bibnamefont
  {Ros{\'e}n}}, \bibinfo {author} {\bibfnamefont {B.~S.}\ \bibnamefont
  {Hsiao}},\ and\ \bibinfo {author} {\bibfnamefont {L.~D.}\ \bibnamefont
  {S{\"o}derberg}},\ }\bibfield  {title} {\bibinfo {title} {Elucidating the
  opportunities and challenges for nanocellulose spinning},\ }\href@noop {}
  {\bibfield  {journal} {\bibinfo  {journal} {Advanced Materials}\ ,\ \bibinfo
  {pages} {2001238}} (\bibinfo {year} {2020})}\BibitemShut {NoStop}%
\bibitem [{\citenamefont {Huang}\ \emph {et~al.}(1991)\citenamefont {Huang},
  \citenamefont {Swanson}, \citenamefont {Lin}, \citenamefont {Schuman},
  \citenamefont {Stinson}, \citenamefont {Chang}, \citenamefont {Hee},
  \citenamefont {Flotte}, \citenamefont {Gregory},\ and\ \citenamefont
  {Puliafito}}]{huang1991optical}%
  \BibitemOpen
  \bibfield  {author} {\bibinfo {author} {\bibfnamefont {D.}~\bibnamefont
  {Huang}}, \bibinfo {author} {\bibfnamefont {E.}~\bibnamefont {Swanson}},
  \bibinfo {author} {\bibfnamefont {C.}~\bibnamefont {Lin}}, \bibinfo {author}
  {\bibfnamefont {J.}~\bibnamefont {Schuman}}, \bibinfo {author} {\bibfnamefont
  {W.}~\bibnamefont {Stinson}}, \bibinfo {author} {\bibfnamefont
  {W.}~\bibnamefont {Chang}}, \bibinfo {author} {\bibfnamefont
  {M.}~\bibnamefont {Hee}}, \bibinfo {author} {\bibfnamefont {T.}~\bibnamefont
  {Flotte}}, \bibinfo {author} {\bibfnamefont {K.}~\bibnamefont {Gregory}},\
  and\ \bibinfo {author} {\bibfnamefont {C.~A.}\ \bibnamefont {Puliafito}},\
  }\bibfield  {title} {\bibinfo {title} {Optical coherence tomography},\
  }\href@noop {} {\bibfield  {journal} {\bibinfo  {journal} {Science}\ }\textbf
  {\bibinfo {volume} {254}},\ \bibinfo {pages} {1178} (\bibinfo {year}
  {1991})}\BibitemShut {NoStop}%
\bibitem [{\citenamefont {Drexler}\ and\ \citenamefont
  {Fujimoto}(2008)}]{drexler2008optical}%
  \BibitemOpen
  \bibfield  {author} {\bibinfo {author} {\bibfnamefont {W.}~\bibnamefont
  {Drexler}}\ and\ \bibinfo {author} {\bibfnamefont {J.~G.}\ \bibnamefont
  {Fujimoto}},\ }\href@noop {} {\emph {\bibinfo {title} {Optical coherence
  tomography: technology and applications}}}\ (\bibinfo  {publisher} {Springer
  Science \& Business Media},\ \bibinfo {year} {2008})\BibitemShut {NoStop}%
\bibitem [{\citenamefont {Leitgeb}\ \emph {et~al.}(2014)\citenamefont
  {Leitgeb}, \citenamefont {Werkmeister}, \citenamefont {Blatter},\ and\
  \citenamefont {Schmetterer}}]{leitgeb2014doppler}%
  \BibitemOpen
  \bibfield  {author} {\bibinfo {author} {\bibfnamefont {R.~A.}\ \bibnamefont
  {Leitgeb}}, \bibinfo {author} {\bibfnamefont {R.~M.}\ \bibnamefont
  {Werkmeister}}, \bibinfo {author} {\bibfnamefont {C.}~\bibnamefont
  {Blatter}},\ and\ \bibinfo {author} {\bibfnamefont {L.}~\bibnamefont
  {Schmetterer}},\ }\bibfield  {title} {\bibinfo {title} {Doppler optical
  coherence tomography},\ }\href@noop {} {\bibfield  {journal} {\bibinfo
  {journal} {Progress in Retinal and Eye Research}\ }\textbf {\bibinfo {volume}
  {41}},\ \bibinfo {pages} {26} (\bibinfo {year} {2014})}\BibitemShut {NoStop}%
\bibitem [{\citenamefont {Roenby}\ \emph {et~al.}(2016)\citenamefont {Roenby},
  \citenamefont {Bredmose},\ and\ \citenamefont
  {Jasak}}]{roenby2016computational}%
  \BibitemOpen
  \bibfield  {author} {\bibinfo {author} {\bibfnamefont {J.}~\bibnamefont
  {Roenby}}, \bibinfo {author} {\bibfnamefont {H.}~\bibnamefont {Bredmose}},\
  and\ \bibinfo {author} {\bibfnamefont {H.}~\bibnamefont {Jasak}},\ }\bibfield
   {title} {\bibinfo {title} {A computational method for sharp interface
  advection},\ }\href@noop {} {\bibfield  {journal} {\bibinfo  {journal} {R.
  Soc. Open Sci.}\ }\textbf {\bibinfo {volume} {3}},\ \bibinfo {pages} {160405}
  (\bibinfo {year} {2016})}\BibitemShut {NoStop}%
\bibitem [{\citenamefont {Brackbill}\ \emph {et~al.}(1992)\citenamefont
  {Brackbill}, \citenamefont {Kothe},\ and\ \citenamefont
  {Zemach}}]{Brackbill1992}%
  \BibitemOpen
  \bibfield  {author} {\bibinfo {author} {\bibfnamefont {J.~U.}\ \bibnamefont
  {Brackbill}}, \bibinfo {author} {\bibfnamefont {D.~B.}\ \bibnamefont
  {Kothe}},\ and\ \bibinfo {author} {\bibfnamefont {C.}~\bibnamefont
  {Zemach}},\ }\bibfield  {title} {\bibinfo {title} {A continuum method for
  modeling surface tension},\ }\href@noop {} {\bibfield  {journal} {\bibinfo
  {journal} {J. Comput. Phys.}\ }\textbf {\bibinfo {volume} {100}},\ \bibinfo
  {pages} {335} (\bibinfo {year} {1992})}\BibitemShut {NoStop}%
\bibitem [{\citenamefont {Cubaud}\ and\ \citenamefont
  {Mason}(2008)}]{Cubaud_2008}%
  \BibitemOpen
  \bibfield  {author} {\bibinfo {author} {\bibfnamefont {T.}~\bibnamefont
  {Cubaud}}\ and\ \bibinfo {author} {\bibfnamefont {T.~G.}\ \bibnamefont
  {Mason}},\ }\bibfield  {title} {\bibinfo {title} {Capillary threads and
  viscous droplets in square microchannels},\ }\href@noop {} {\bibfield
  {journal} {\bibinfo  {journal} {Phys. Fluids}\ }\textbf {\bibinfo {volume}
  {20}},\ \bibinfo {pages} {053302} (\bibinfo {year} {2008})}\BibitemShut
  {NoStop}%
\bibitem [{\citenamefont {Jeffery}(1922)}]{jeffery1922}%
  \BibitemOpen
  \bibfield  {author} {\bibinfo {author} {\bibfnamefont {G.~B.}\ \bibnamefont
  {Jeffery}},\ }\bibfield  {title} {\bibinfo {title} {The motion of ellipsoidal
  particles immersed in a viscous fluid},\ }\href@noop {} {\bibfield  {journal}
  {\bibinfo  {journal} {Proceedings of the Royal Society of London. Series A,
  Containing papers of a mathematical and physical character}\ }\textbf
  {\bibinfo {volume} {102}},\ \bibinfo {pages} {161} (\bibinfo {year}
  {1922})}\BibitemShut {NoStop}%
\bibitem [{\citenamefont {H{\aa}kansson}\ \emph {et~al.}(2016)\citenamefont
  {H{\aa}kansson}, \citenamefont {Lundell}, \citenamefont {Prahl-Wittberg},\
  and\ \citenamefont {S{\"o}derberg}}]{H_kansson_2016}%
  \BibitemOpen
  \bibfield  {author} {\bibinfo {author} {\bibfnamefont {K.~M.~O.}\
  \bibnamefont {H{\aa}kansson}}, \bibinfo {author} {\bibfnamefont
  {F.}~\bibnamefont {Lundell}}, \bibinfo {author} {\bibfnamefont
  {L.}~\bibnamefont {Prahl-Wittberg}},\ and\ \bibinfo {author} {\bibfnamefont
  {L.~D.}\ \bibnamefont {S{\"o}derberg}},\ }\bibfield  {title} {\bibinfo
  {title} {Nanofibril alignment in flow focusing: Measurements and
  calculations},\ }\href@noop {} {\bibfield  {journal} {\bibinfo  {journal} {J.
  Phys. Chem. B}\ }\textbf {\bibinfo {volume} {120}},\ \bibinfo {pages} {6674}
  (\bibinfo {year} {2016})}\BibitemShut {NoStop}%
\end{thebibliography}%

\end{document}